\documentclass[iicol]{sn-jnl}

\usepackage{lscape}
 \usepackage{natbib}
 \usepackage{xcolor}
 \usepackage{float}

\jyear{2023}%

\theoremstyle{thmstyleone}%
\theoremstyle{thmstyletwo}%

\theoremstyle{thmstylethree}%

\definecolor{gray2}{rgb}{221,221,221}
 \definecolor{lightgray2}{rgb}{229, 228, 226}
\begin{document}

\title[Directional density-based clustering]{Directional density-based clustering}


\author*[1]{\fnm{Paula} \sur{Saavedra-Nieves}}\email{paula.saavedra@usc.es}

\author[2]{\fnm{Mart\'in} \sur{Fern\'andez-P\'erez}}
\email{martin.fernandez.perez0@rai.usc.es}
\equalcont{These authors contributed equally to this work.}


\affil*[1]{\orgdiv{CITMAga}, \orgname{Universidade de Santiago de Compostela}, \orgaddress{\street{Facultade de Matemáticas}, \city{Santiago de Compostela}, \postcode{15705}, \state{Galicia}, \country{Spain}}}

\affil[2]{\orgname{Universidade de Santiago de Compostela}}



\abstract{Density-based clustering methodology has been widely considered in the statistical literature for classifying Euclidean observations. However, this approach has not been contemplated for directional data yet. In this work, directional density-based clustering methodology is fully established for the unit hypersphere by solving the computational problems associated to high dimensional spaces. We also provide a circular and spherical exploratory tool for studying the effect of the smoothing parameter when kernel density estimation methods are considered. An extensive simulation study shows the performance of the resulting classification procedure for the circle and for the sphere. The methodology is also applied to analyse an exoplanets dataset.}

\keywords{cCluster, directional clustering, kernel density estimation, sCluster}



\maketitle

\section{Introduction}\label{sec1}

Clustering for directional data has achieved a considerable relevance over the last decades, specially amongst the machine learning community. \cite{pewsey2021recent} offer a brief but also a complete revision on this topic. The most popular approaches for directional clustering are spherical $\kappa-$means with cosine similarity (see \citealp{dhillon2001concept}) and the use of (finite) mixture models with von Mises-Fisher components (see \citealp{banerjee2005clustering}). However, there exist more robust alternatives in the literature
that do not require the specification of the number of groups in advance. This is the case of the algorithm introduced in \cite{hung2015intuitive} but also of modal clustering where the notions of cluster and mode are associated. Modal algorithms apply mode-seeking numerical methods and assigning the same cluster to those data that are iteratively shifted to the same limit value (see \citealp{oba2005multi}).

 \begin{figure*}
	\begin{picture}(-210,400)
	   \put(-65,195){\includegraphics[scale=.5]{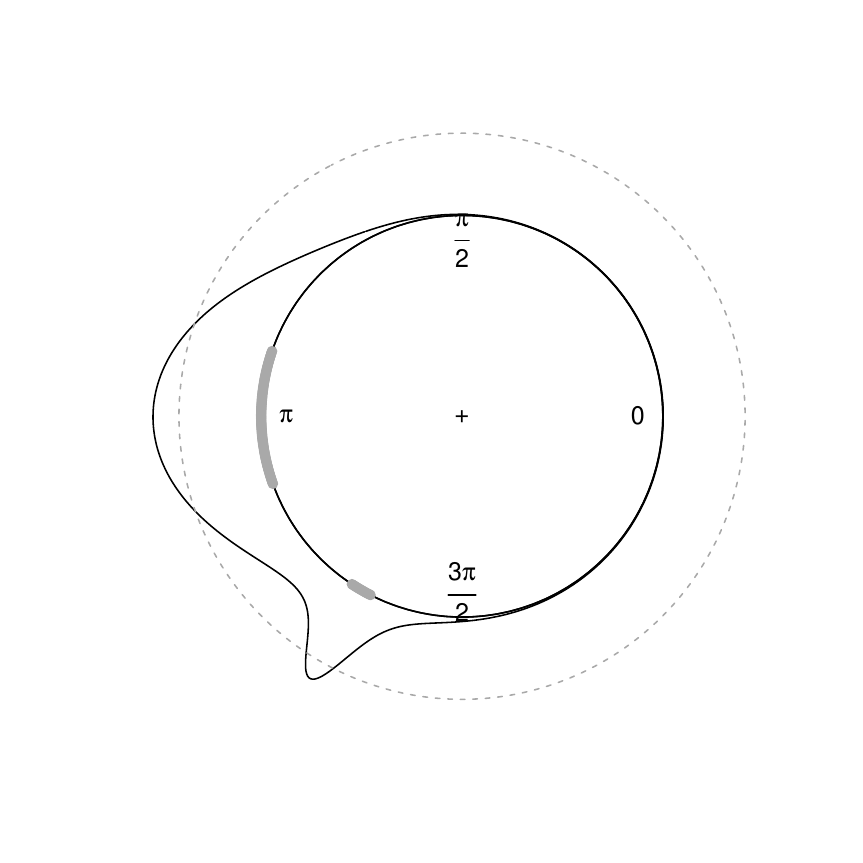}}
	   \put(109,200){\includegraphics[scale=.5]{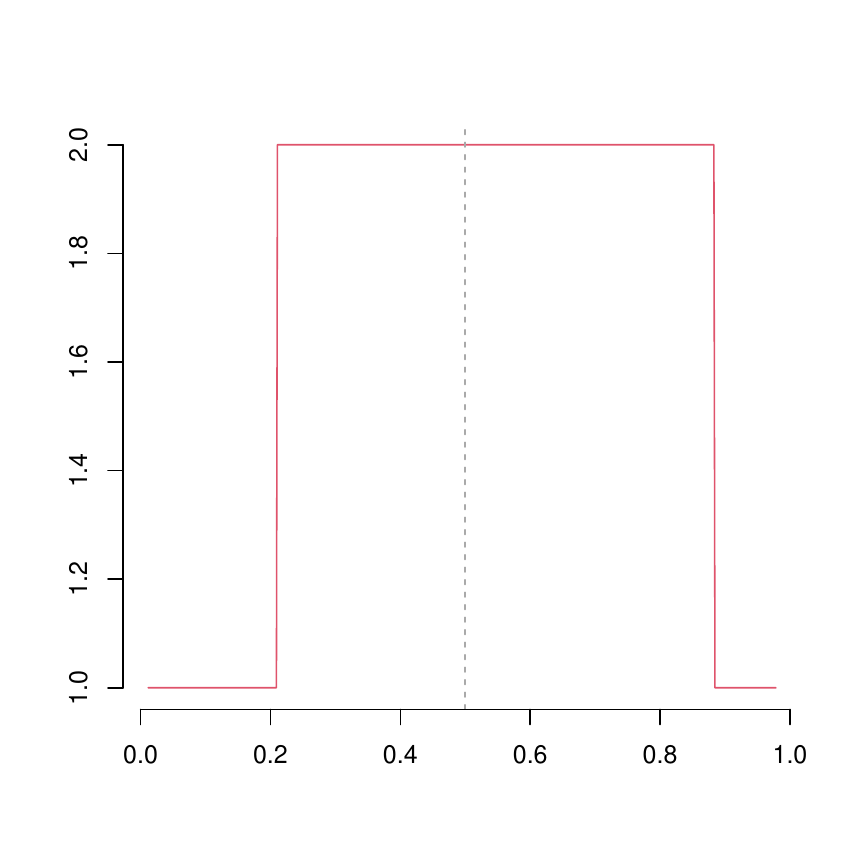}}
	   \put(315,200){\includegraphics[scale=.5]{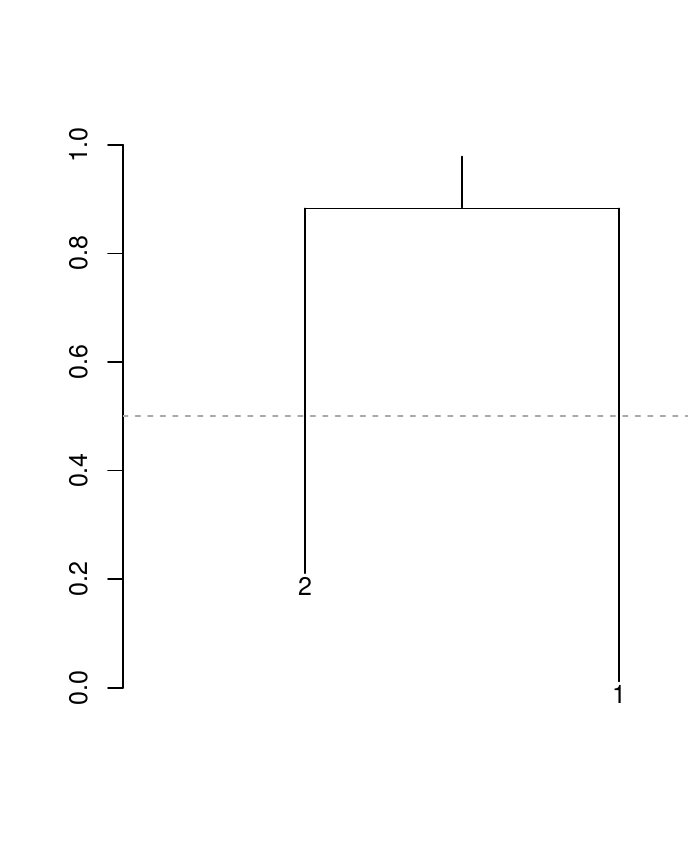}}
    \end{picture}  
    \vspace{-7.3cm}
	\caption{Circular density function and $L_f(\tau)$ for $\tau=0.5$ (left) and corresponding modes function (center) and cluster tree (right). Dotted lines on the modes function and on the cluster tree correspond to $\tau=0.5$.}\label{fig1:densidade_teorica_circular}
\end{figure*}
 		
The connection between clusters and modes is also present in density-based clustering approach so far just developed for Euclidean data (see \citealp{chacon2015population} for a deeper understanding of the relationship between both concepts). Under this perspective, clusters may be thought of as high density regions separated from other such regions by low density areas. In fact, for any $t> 0$, \cite{hartigan1975clustering} puts forward the idea of connecting the notion of clusters with the connected components of the level set
\begin{equation}\label{lgt}
	G_g(t)=\{x\in \mathbb{R}^d:g(x)\geq t\}
\end{equation} 
 where $g$ denotes the density function of a $\mathbb{R}^d-$valued random vector $Y$. One non-minor practical problem of the definition in (\ref{lgt}) is that it relies on the user-specified level $t$ and which poses some drawbacks for clustering interpretation. The observations belonging to each cluster or, equivalently, to each connected component, should be related to a probability content (depending on the objectives of the clustering analysis) instead of a threshold of the level set. The definition of highest density regions (HDR) solves this problem conveniently. Given $\tau\in(0,1)$, the $100(1 - \tau)$\% HDR is the subset \begin{equation}\label{lgggt8}
 L_g(\tau)=\{x\in \mathbb{R}^d:g(x)\geq g_\tau\}
 \end{equation}where $g_\tau$ can be seen as the largest constant such that 
$$\mathbb{P}(Y\in L_g(\tau))\geq 1-\tau$$
with respect to the distribution induced by $g$. For small values of $\tau$, $L_g(\tau)$ is almost equal to the support of the distribution. However, for large values of $\tau$, $L_g(\tau)$ is equal to the greatest modes and, therefore, the most differentiated clusters can be easily identified.

This cluster formulation has been widely studied in the literature for Euclidean data perhaps because it avoids several shortcomings of other clustering methods such as slow convergence and the specification of the number of groups, of initial partitions or of stopping rules. Concretely, \cite{azzalini2007clustering} establish the definitions of the empirical mode function and the cluster tree from the hierarchical structure generated by connected components of level sets introduced in (\ref{lgt}) and (\ref{lgggt8}) by partially addressing the computational problem of their computation. A suitable modification of the Silhouette information is presented in \cite{menardi2011density}. It aims at evaluating the quality of clusters under this approach. Aspects related to computational complexity of determining the connected components are fully solved in \cite{stuetzle2010generalized} and \cite{menardi2014advancement}.

 The main goal of this work is to generalise density-based clustering techniques in \cite{stuetzle2010generalized} and \cite{menardi2014advancement} for data supported on the unit hypersphere. The first step is to establish the definition of cluster in \cite{hartigan1975clustering} for directional data. Recently, \cite{saavedra2022nonparametric} generalise the definition of sets in equations (\ref{lgt}) and (\ref{lgggt8}) in this setting. Specifically, given a random vector $X$ taking values on a $d$-dimensional unit sphere $S^{d-1}$ with density $f$ and a level $t>0$, the directional level set is defined as
 \begin{equation}\label{G(t)}
 G_f(t)=\{x\in S^{d-1}:f(x)\geq t\}.
 \end{equation}
 As in the Euclidean setting, the level $t$ is usually unknown and, for practical purposes. Therefore, \cite{saavedra2022nonparametric} also extended the concept of HDRs.  Given $\tau\in(0,1)$, the $100(1 - \tau)$\% HDR is the subset
 	\begin{equation}\label{conjuntonivel2}
 	L_f(\tau)=\{x\in S^{d-1}:f(x)\geq f_\tau\}
 	\end{equation}where $f_\tau$ can be seen as the largest constant such that 
$$\mathbb{P}(X\in L_f(\tau))\geq 1-\tau$$with respect to the distribution induced by $f$. As an illustration, Figure \ref{fig1:densidade_teorica_circular} (left) shows a HDR when $\tau=0.5$ for the represented circular density function. As before, if large values of $\tau$ are considered, $L_f(\tau)$ is equal to the greatest modes. However, for small values of $\tau$, it is almost equal to the support of the distribution. 
 
Directional cluster definition via
connected components as in \cite{hartigan1975clustering} is straightforward from Equations (\ref{G(t)}) and (\ref{conjuntonivel2}). Establishing the population and empirical versions of the directional mode function and cluster tree, are also completely natural tasks to generalise this methodology. However, the main drawback to define them is the computational problem derived from the computation of the connected components of the empirical HDRs, specially in high-dimensional spaces. In this work, we will propose a novel algorithm for determining the connected components of directional HDRs on the unit hypersphere. Therefore, directional density-based clustering will became a viable methodology in arbitrary dimension. Additionally, a exploratory tool will be provided for clustering analysis on the unit circle and sphere.

This work is organised as follows. Section \ref{sec2} generalises density-based clustering methods for directional data on the unit hypersphere. Specifically, directional mode function and the associated cluster tree are introduced in Section \ref{sec21}. Section \ref{sec22} presents their empirical versions from kernel density based methods. An algorithm for computing the directional connected components is introduced. Furthermore, an exploratory tool for circular and spherical data is also developed in Section \ref{sec:explo}. It allows to analyse the influence of the smoothing parameter for density-based clustering methods. Following \cite{azzalini2007clustering}, a classification procedure derived from this directional clustering approach is presented in Section \ref{sec:classification}. Its practical performance is checked through an extensive simulation study in Section \ref{Sim}. The effect of considering different bandwidths parameters is analysed. Finally, this methodology is applied for grouping data on exoplanets in Section \ref{realdata}. 

\section{Directional modes and clusters}\label{sec2}

Density-based clustering techniques are extended to the directional setting next. Although this methodology generalisation is not entirely straightforward, the resulting outputs are very similar to those obtained in \cite{azzalini2007clustering} which is an interpretation advantage.

\subsection{Mode function and cluster tree}\label{sec21}
Given $\tau\in(0,1)$ and the density function $f$ (in what follows, differentiable everywhere), the HDR $L_f(\tau)\subset S^{d-1}$ may be a connected set or not. Figure \ref{fig1:densidade_teorica_circular} (left) illustrates this idea for a circular density where threshold $f_{0.5}$, represented through a dotted grey line, leads to two connected sets. Obviously, 
the number of connected components varies with $\tau$ and the evaluation of this number is more difficult as $d$ increases. Therefore, there is a
correspondence between the probability content $1-\tau$ of HDRs and the associated number of components of $L_f(\tau)$. This allows to define the mode function  $m$, a step function which assigns the number of connected components of $L_f(\tau)$ to the probability content $1-\tau$ as
varies in $(0, 1)$. For $\tau = 0$ and $\tau = 1$, we define $m(0)=m(1)= 0$. Figure \ref{fig1:densidade_teorica_circular} (center) shows the function $m$ corresponding to the density represented on the left. Remark that values of $1-\tau$ are represented on x-axis and y-axis contains the corresponding number of connected components.

Following \cite{azzalini2007clustering}, increments of function $m$ correspond to the appearance of one or more modes of the directional density $f$, whereas the decrements correspond to the fusion of two or more groups associated with
existing modes. As $f_{\tau}$ varies, the connected components of
$L_f(\tau)$ generate a hierarchical structure which may be represented
in the form of a tree (see \citealp{hartigan1975clustering} and \citealp{stuetzle2003estimating} for further details). Figure \ref{fig1:densidade_teorica_circular} (right) shows this tree plot for the density represented on the left. Again, values of the probability content $1-\tau$ are represented on y-axis.

\begin{figure*} 
    \begin{picture}(0,240)
    \put(125,65){\includegraphics[scale=0.7]{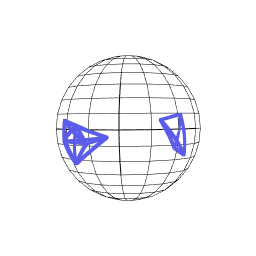}}
        \put(-40,65){\includegraphics[scale=0.7]{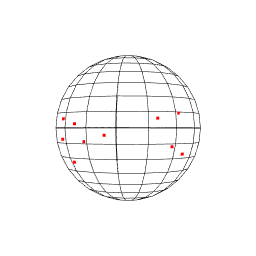}}
         \put(289,65){\includegraphics[scale=0.7]{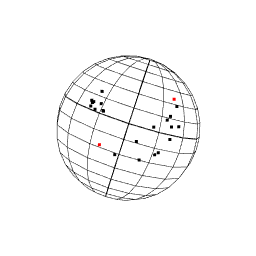}}
        \end{picture} \vspace{-3.5cm}\\
    \caption{For a certain $k>0$, $S(k)$ (left) and $\gamma_i^j$ such that $w_i^j\geq k$ (center). \emph{Lump and banana} example on the sphere (right).}
    \label{fig:algor}
\end{figure*}  
\subsection{Empirical mode function and cluster tree}\label{sec22}

The empirical version of the HDR established in (\ref{conjuntonivel2}) can be computed from plug-in methods. This estimation procedure is the most common choice for reconstructing density level sets in the directional space from a nonparametric approach (see \citealp{saavedra2022nonparametric}, \citealp{cholaquidis2022level} or \citealp{cuevas2006plug}). Given a random sample $\mathcal{X}_n=\{X_1,\cdots,X_n\}\in S^{d-1}$ of the unknown directional density $f$, the level set $L_f(\tau)$ can be reconstructed as
\begin{equation}\label{Gtest}
\hat{L}_f(\tau)=\{x\in S^{d-1}:f_n(x)\geq \hat{f}_{\tau}\}
\end{equation}
where $f_n$ denotes a nonparametric directional density estimator and $\hat{f}_{\tau}$, a threshold estimator. In principle, the estimator in (\ref{Gtest}) is not linked to any specific method for density estimation. The only restriction required in this work is that $f_n(X_i) <\infty$ for $i = 1, \cdots , n$. Therefore, we will consider the kernel estimator on $S^{d-1}$ provided in \cite{bai1989kernel} ($d>2$) (for further details, see also \citealp{hall1987kernel} and \citealp{klemela2000estimation}). Given $\mathcal{X}_n$, this directional kernel density estimator at a point $x\in S^{d-1}$ is defined as
\begin{equation}
\label{estimacionnucleo}
f_n(x)= \frac{1}{n}  \sum_{i=1}^n K_{vM}(x;X_i;1/h^2),
\end{equation}where $1/h^2 > 0$ is concentration parameter and $K_{vM}$ usually corresponds to the von Mises-Fisher kernel density.

  \begin{figure*}
	\begin{picture}(-200,410)
	\put(-70,195){\includegraphics[scale=.45]{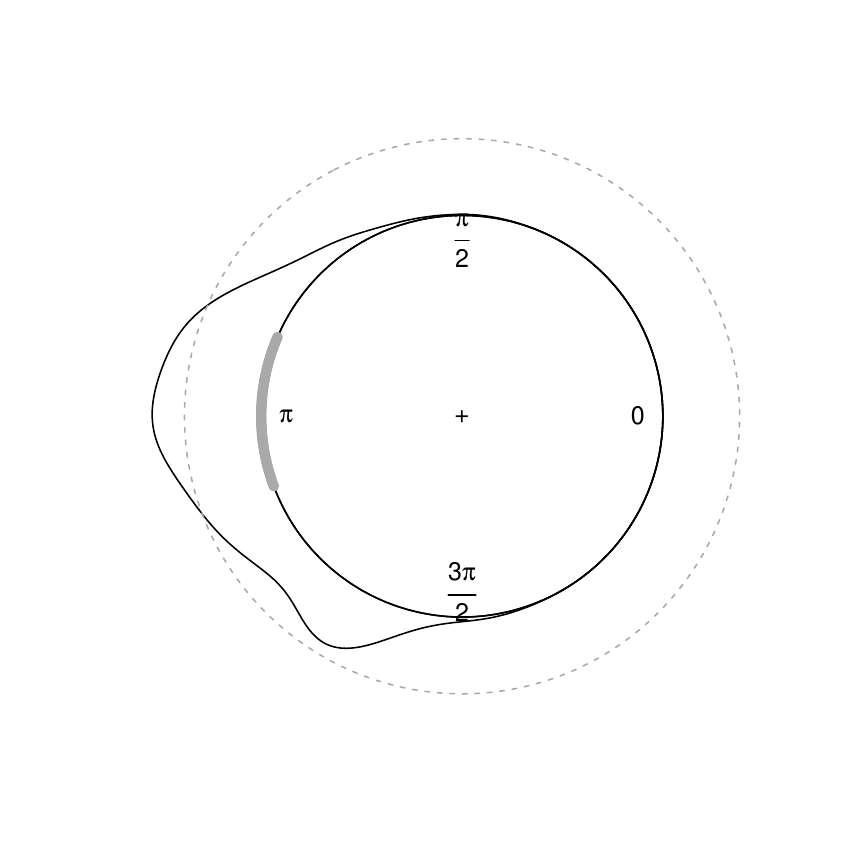}}
	\put(88,200){\includegraphics[scale=.45]{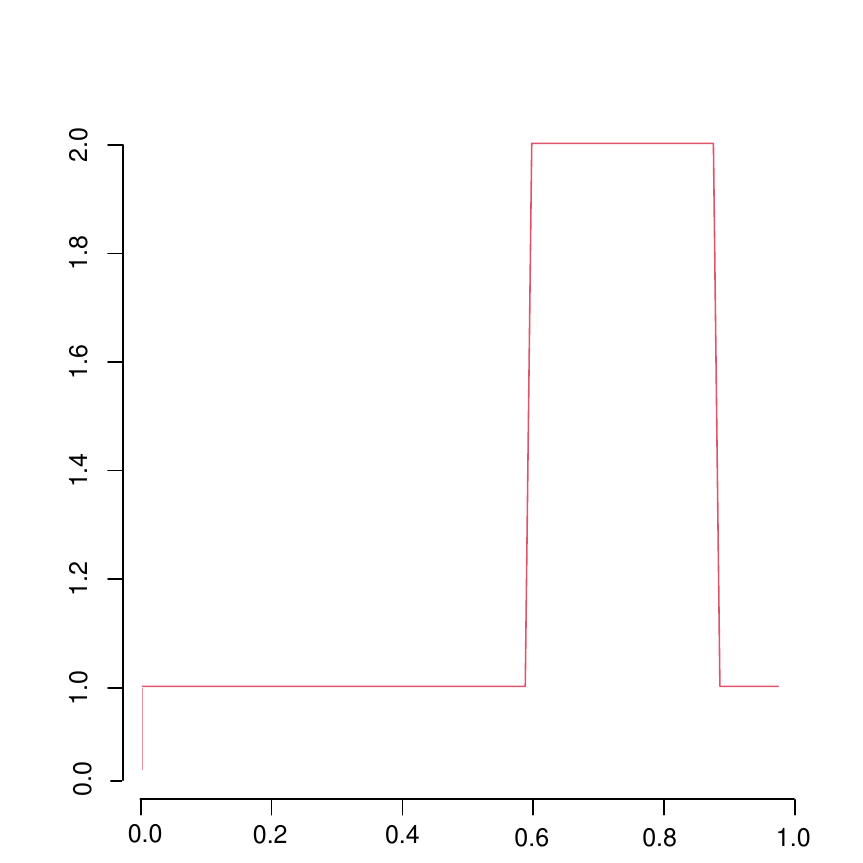}}
		\put(290,168){\includegraphics[scale=.51]{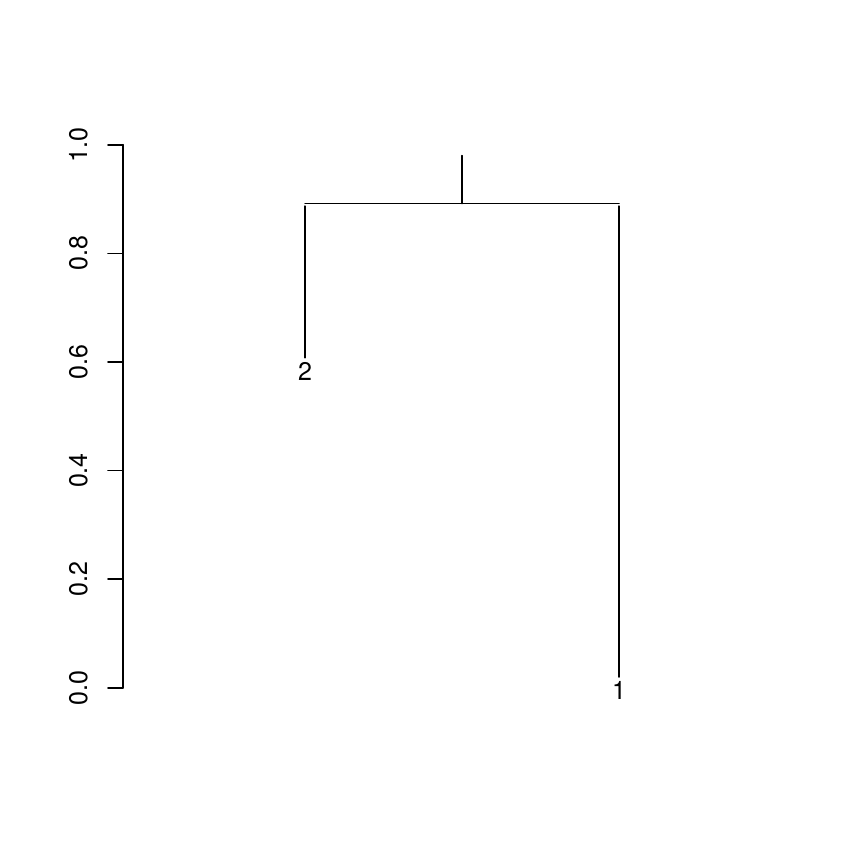}}		
		\put(-53,15){\includegraphics[scale=.64]{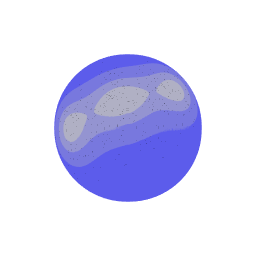}}
	\put(95,0){\includegraphics[scale=.4]{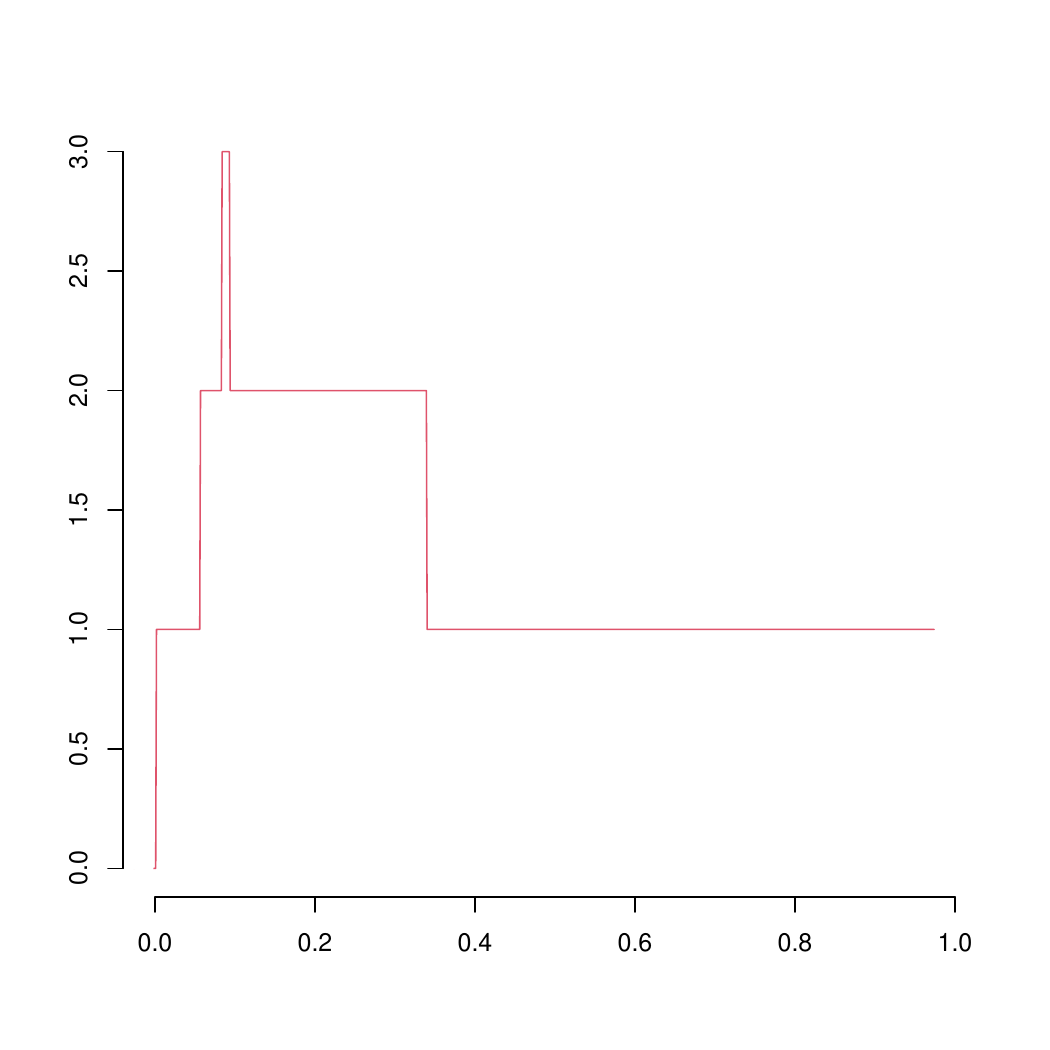}}
		\put(300,-8){\includegraphics[scale=.45]{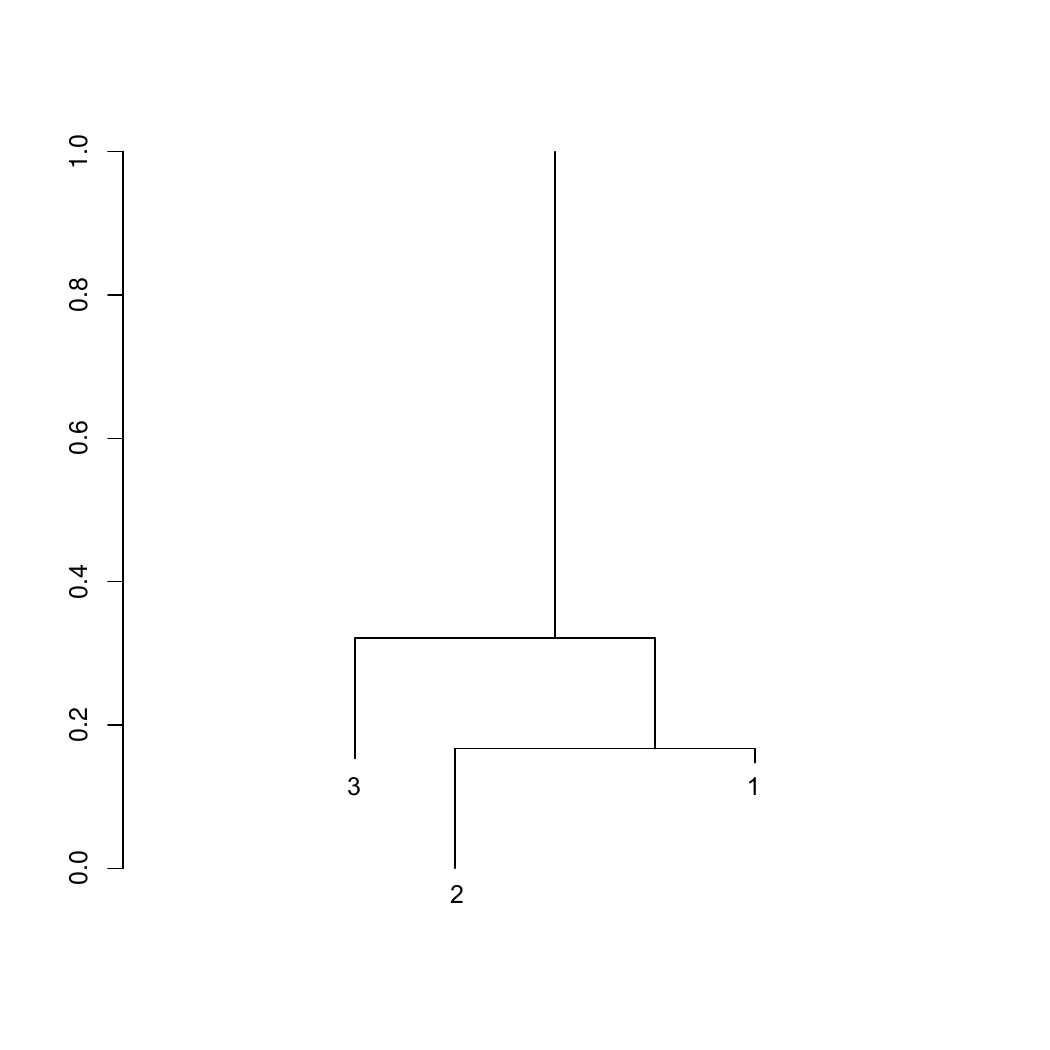}}			
\end{picture}  \vspace{-0.3cm}
	\caption{In the first row, circular kernel density estimation and $\hat{L}_f(\tau)$ (gray color) computed from a sample $\mathcal{X}_{500}$ of the circular density represented in Figure \ref{fig1:densidade_teorica_circular} (left), corresponding empirical modes function (center) and associated cluster tree (right). In the second row, spherical density function and sample  $\mathcal{X}_{1000}$ (left), corresponding empirical modes function (center) and cluster tree (right) computed from $\mathcal{X}_{1000}$.}\label{fig2:estimado}
\end{figure*}
Note that the kernel estimator in (\ref{estimacionnucleo}) can be viewed as a mixture of von Mises-Fisher. Furthermore, the concentration parameter $1/h^2$ plays an analogous role to the bandwidth in the Euclidean case. For small values of $1/h^2$, the density estimator is oversmoothed and  spurious modes are avoided. The opposite effect is obtained as $1/h^2$ increases: with a large value of $1/h^2$, the estimator is clearly undersmoothing the underlying target density. Hence, the choice of the bandwidth parameter $h$ is a crucial issue that has been already considered in the directional literature. We defer until later discussion of the choice of the smoothing parameter for the clustering approach introduced in this work. As for the threshold $f_{\tau}$, it could be estimated as the $\tau-$quantile of the empirical distribution of $\{f_n(X_1),\cdots,f_n(X_n)\}$ (see \citealp{hyndman1996computing} for more details).

To find the empirical analogue of the mode function $m$,
we must establish a procedure to determine the connected components of $\hat{L}_f(\tau)$. For $d-$dimensional Euclidean observations, \cite{azzalini2007clustering} use the Delaunay
triangulation in order to detect the connected
components. However, its computational complexity grows
exponentially with the dimensionality of data thus making the triangulation unfeasible for high dimensions. \cite{stuetzle2010generalized} build a weighted graph with edges associated to the minimum value of the density function along the segments joining pairs of sample observations. Then, the subgraph consisting of the edges and vertices with weights bigger than $\hat{f}_{\tau}$ is selected. Remark that two sample points in the same connected component of this subgraph are guaranteed to lie in the same connected component of $\hat{L}_{f}(\tau)$. A similar perspective is established in \cite{menardi2014advancement}. In this case, two euclidean observations are assumed to be connected if the density function, evaluated along the segment joining them, does not present any valley of considerable extent. 
 
Following the procedure in \cite{stuetzle2010generalized}, Algorithm \ref{algo1} contains a novel proposal for solving the problem of identifying the connected components of a HDR on the unit hypersphere. A weighted graph is also constructed with sample points as vertices. However, the weights of edges are determined from the minimum value of the density evaluation on the geodesic curve between two sample points. Although it has not been formalized in Algorithm \ref{algo1}, negligible density valleys could be also avoided by generalising the  definition of the index established in \cite{menardi2014advancement} for 
quantifying the valley sizes. Remark that Algorithm \ref{algo1} also mishandles some non-convex situations. The example known as \emph{lump and banana} presented in \cite{stuetzle2010generalized} is adapted to the sphere in Figure \ref{fig:algor} (right). Red observations lying in the upper and the lower half of the \emph{banana} belong to
the same cluster. However, there is no edge in $\mathcal{G}_k$ connecting them. Even so, these observations are still allocated to the same cluster because they will turn out to be connected by a sequence of edges linking pairwise-connected points.

\begin{algorithm}
\caption{Main steps of directional density-based clustering algorithm.}\label{algo1}
\begin{algorithmic} 
\State Compute $f_n$. 
\State Initialize the graph $\mathcal{G}$ with vertices $\mathcal{X}_n$ and no edges. 
\ForAll{$X_i,X_j\in \mathcal{X}_n$, $i\neq j$} 
\State Evaluate $f_n$ along $\gamma_{i}^j$, the geodesic between $X_i$ and $X_j$.
\State Add edge $(X_i, X_j ) \in \mathcal{G}$ with weight $w_{i}^j$ equal to the minimum value of $f_n(\gamma_{i}^j)$.
\EndFor
\While{$ 0 \leq k \leq \max{f_n}$}
\State Identify $S(k) = \{X_i \in\mathcal{G} : w_{i}^j\geq k \mbox{ or, simply, } f_n(X_i)\geq k \}$.
\State Identify $E(k)=\{(X_i, X_j )\in\mathcal{G}:w_{i}^j\geq k\}$.
\State Extract from $\mathcal{G}$ the subgraph $\mathcal{G}_k$, formed by vertices
in $S(k)$ and edges in $E(k)$.
\State Find the graph connected components of $\mathcal{G}_k$ (e.g. by
depth-first-search).
\State next $k$.
\EndWhile
\State Build the cluster tree.
\end{algorithmic}
\end{algorithm}

The empirical mode function $\hat{m}$ can be obtained from 
Algorithm \ref{algo1}. Remark that it must be performed for a range of values of $k$, $0\leq k\leq \max{f_n}$. This grid can be defined from the selection of a set of equally spaced values of $\tau$ ($0 < \tau < 1$). Then, estimations of $\hat{f}_{\tau}$ for each considered $\tau$, will determine the range of values for $k$ to be considered. Figure \ref{fig2:estimado} shows a circular (first row) a spherical (second row) empirical mode functions (center) obtained from Algorithm \ref{algo1}. 

Increments of mode function $\hat{m}$ as $1-\tau$ ranges from $0$ to $1$ correspond to the appearance of new clusters and, vice versa, the decrements correspond to the merging of clusters. Specifically, a value of $1-\tau_1$ corresponding to an increment of $\hat{m}$ denotes the \emph{birth} of as many clusters as the increment of $\hat{m}$, and the sample points comprising these clusters may be identified. Similarly, if $1-\tau_2$ is a value where $\hat{m}$ decreases, two or more clusters are merging. In this case, comparison of the sample points in $\hat{L}_f(\tau_2)$ and $\hat{L}_f(\tau_3)$ with $\tau_3>\tau_2$ allows us to detect which groups are merging at this level. Proceeding sequentially from values of $1-\tau=0$ to $1-\tau= 1$, the whole tree structure
of the clusters is identified. Figure \ref{fig2:estimado} (right) contains the corresponding circular and spherical cluster trees by reflecting the original idea
of \cite{hartigan1975clustering}, except that the vertical axis is related to the empirical probability contents $1-\tau$, instead of density level.

\subsection{Cluster exploratory tool for bandwidth selection}\label{sec:explo}
  
 \begin{figure*}\vspace*{-.55cm}   
	\includegraphics[scale=.4]{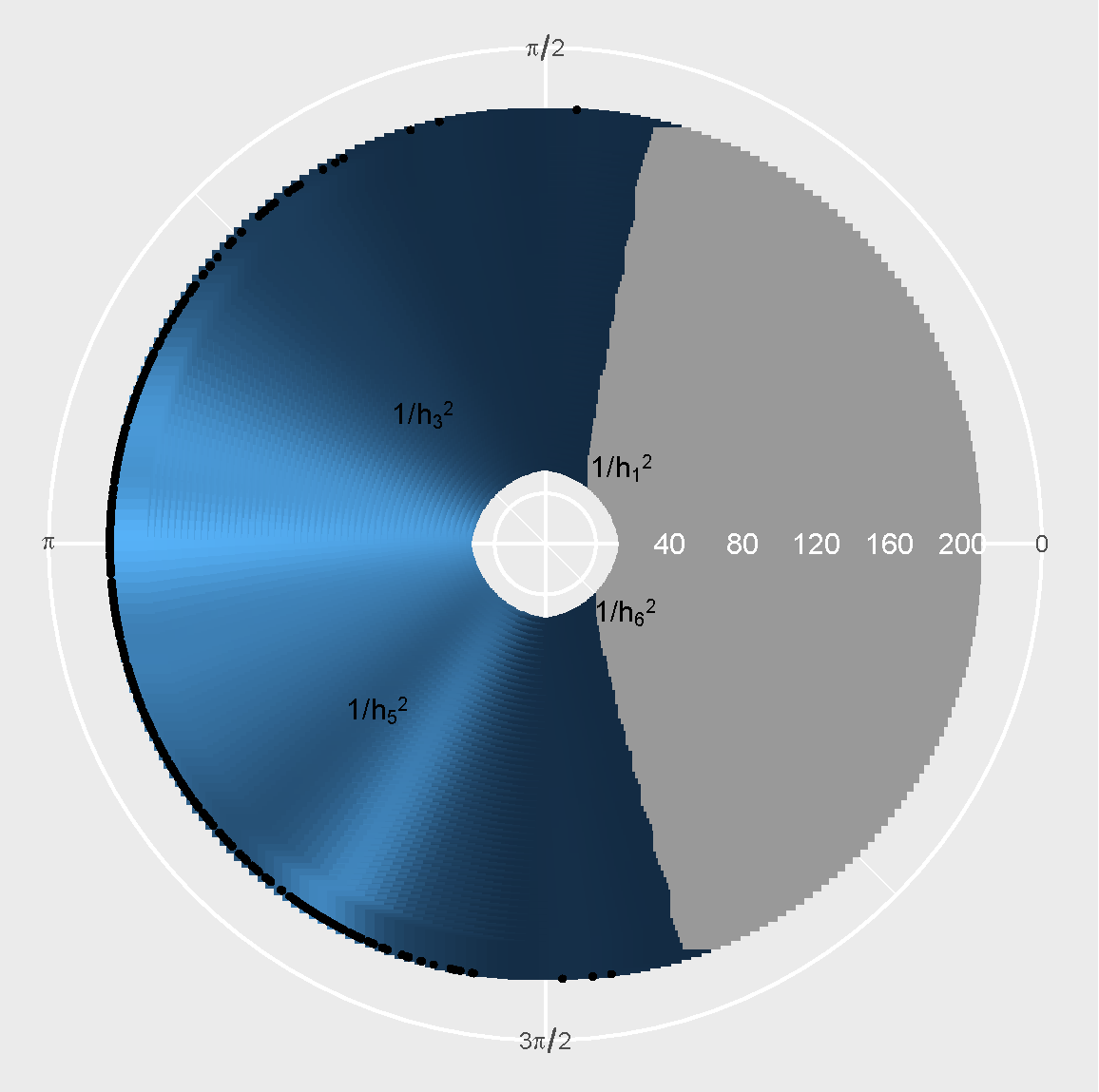}$\hspace*{.3cm}$
	\includegraphics[width=9.7cm,height=6.2cm]{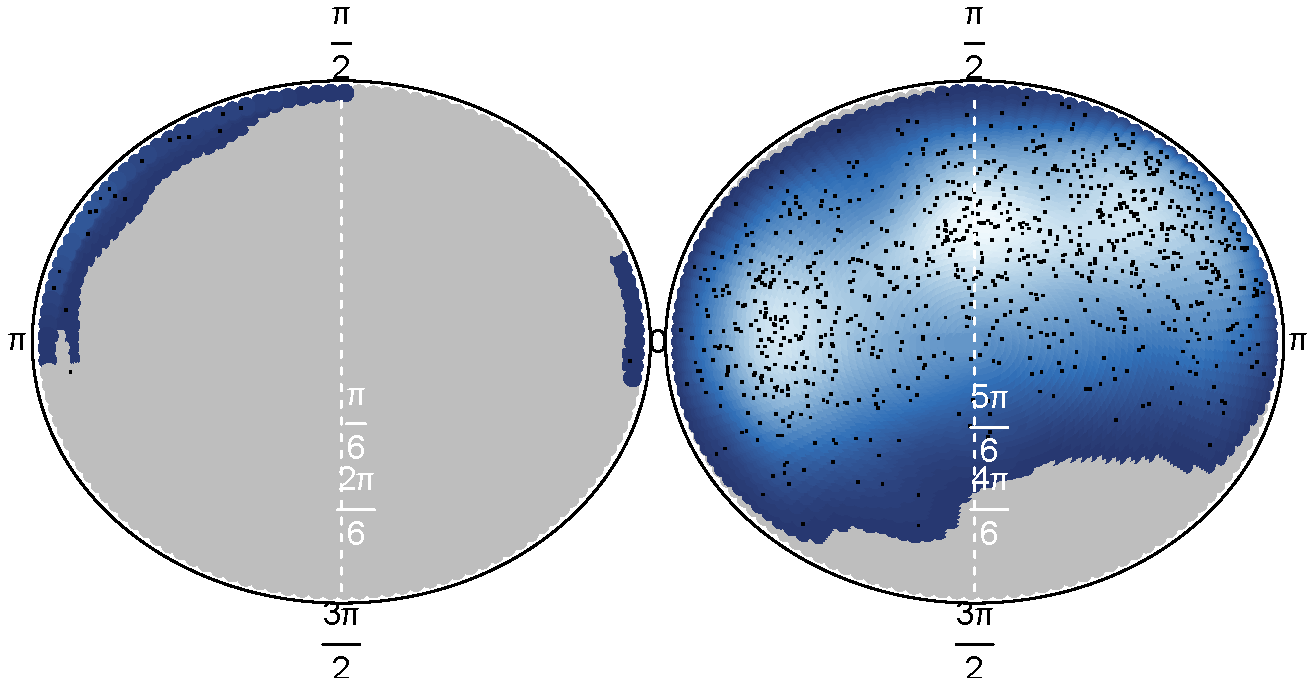} 
	\caption{c\textbf{C}luster (left) and s\textbf{C}luster (right) for the circular and spherical samples already considered in Figure \ref{fig2:estimado}.  }\label{fig:cluster}
\end{figure*}

The clustering method described does not depend on a specific nonparametric density estimator. Among the many possible alternatives, we chose a kernel
method with von Mises-Fisher kernel in this work. The critical issue is the choice of $h$. There is an extensive and specialised literature dealing with this problem also in the directional setting. For instance, \cite{taylor2008automatic} propose a circular rule-of-thumb ($h_1$); an improved version of this selector (namely $h_2$) was presented in \cite{oliveira2013nonparametric}; additionally, classical methods such as cross-validation (likelihood $h_3$ and least squares $h_4$) were introduced by \cite{hall1987kernel}; there also exist bootstrap approaches as in \cite{di2011kernel} ($h_6$) or \cite{saavedra2022nonparametric}); and, alternatively, \cite{garcia2013exact} introduce a rule-of-thumb selector for the unit hypersphere ($h_7$) and both asymptotic and exact mixtures procedures described in Algorithms 1 and 2 ($h_8$ and $h_9$, respectively). 

As we mention before, the consideration of very small values of $h$ allows to identify spurious modes corresponding to inauthentic clusters. However, the role of above selectors in directional density-based clustering has not been analysed yet. Figure \ref{fig:cluster} shows two exploratory tools developed in this work for circular (c\textbf{C}luster) and  spherical (s\textbf{C}luster) data analysis. Their main aim is to study the influence of $h$ on the number of clusters in the circular and spherical settings, respectively. 

Specifically, c\textbf{C}luster (left) represents for a range of values of $1/h^2$ (represented on the positive x-axis), the corresponding values of kernel density estimations (obtained from a specific random sample) on the circle of center the origin and radius $1/h^2$. Different colors intensities are used according to the different values of the threshold $f_{\tau}$. Gray color corresponds values of kernel estimator equal to zero. The rest of density estimations are represented using a blue color scale. Low values of the kernel estimations are represented with dark blue and high values, with light blue. Additionally, sample points for computing the kernel density estimator is represented on the outer circle and values of $1/h_i^2$, $i\in\{1,3,5,6\}$ are also pointed. This graphical representation allows to identify the cluster structure for a specific value of the bandwidth $h$. In particular, the maximum number of clusters estimated for a specific value of $h$ can be identified. For the circular and spherical samples already considered in Figure \ref{fig2:estimado}, it can be noted that the number of clusters/modes is bigger than two when $1/h^2$ exceeds the value 100.

As regards s\textbf{C}luster, it is an animation fully shown in the Supplementary Material that contains a sequence of kernel density estimators corresponding to specific values of the bandwidth in a predefined range. In particular, estimations obtained from classical bandwidths such as $h_3$, $h_4$, $h_7$, $h_8$ and $h_9$ are incorporated. Figure \ref{fig:cluster} (right) represents the kernel density estimation with fixed bandwidth  $h_4=0.13$. Remark that the scheme in \cite{vuollo2018scale} has been imitated by using the same colour scale as in the circular representation. Concretely, the spherical kernel density estimate, for a given smoothing parameter, is visualised by dividing the sphere into two hemispheres and project them separately onto two disks. Note that  sample points considered for kernel estimation are also represented.

 \subsection{Directional density-based classification}\label{sec:classification}
 
Application of Algorithm \ref{algo1} for a range of values of $k$ ($0\leq k\leq \max{f_n}$) allows to obtain the smallest $k$ with the maximum number of clusters $n_c$ detected. Following \cite{azzalini2007clustering}, we define the \emph{cluster cores} as the subsets of points in $\mathcal{X}_n$ that belong to each of these $n_c$ groups. Of course, a proportion of sample points will be outside the cluster cores and, therefore, they are not labelled. Allocation of these unlabelled points to existing groups is essentially a classification problem, although of a rather peculiar type. The unusual aspect is that the unlabelled points are not positioned randomly in $S^{d-1}$, but are inevitably on the outskirts of the $n_c$ existing groups.

There is a wide choice of classification methods. Given an unallocated data point $x_0\in S^{d-1}$, the Euclidean approach in \cite{azzalini2007clustering} suggests the following directional procedure: 
\begin{itemize}
    \item [(1)] Determine the kernel estimated density $f_{n,j}(x_0)$ based on the observations in $\mathcal{X}_n$ already
assigned to group $j$ for all $j = 1, 2,\cdots, n_c$. 
\item[(2)] For each $j = 1, 2,\cdots, n_c$, compute $$r_j (x_0) = \frac{f_{n,j} (x_0)}{\max_{i \neq j} f_{n,i} (x_0)}.$$ 
\item [(3)] Assign
$x_0$ to the group $J\in \{1,\cdots,n_c\}$ verifying that $$r_J(x_0)=\max\{r_j(x_0),\mbox{ }j=1,\cdots,n_c\}.$$  
\end{itemize}

 As detailed in \cite{azzalini2007clustering}, the  implementation of this idea for classifying a set of unallocated points may include sequential density estimates or block allocation methods. We will estimate $n_c$ density functions $f_{n,j}$ (for $j = 1, \cdots, n_c$) once and, then, all unlabelled points will be classified by using these estimates.

 \section{Simulations}\label{Sim}
 Circular and spherical simulations have been run in order to explore the performance of the classification algorithm proposed in Section \ref{sec:classification}. Concretely, it will be compared with the classical $\kappa-$means method implemented in the R package \texttt{skmeans}\footnote{\url{https://CRAN.R-project.org/package=skmeans}} following the approach in \cite{dhillon2002iterative}. The value of $\kappa$ for simulations was fixed by taking the real number of populations involved. The impact of bandwidth selection in the new clustering proposal will be also checked. Specifically, performance of bandwidths $h_1$, $h_2$, $h_3$, $h_4$, $h_5$ and $h_6$ introduced in Section \ref{sec:explo} will be studied in the circular setting. For spherical data, $h_3$, $h_4$, $h_7$, $h_8$ and $h_9$ will be considered. All of them are implemented in the R packages \texttt{NPCirc}\footnote{\url{https://CRAN.R-project.org/package=NPCirc}} and \texttt{DirStats}\footnote{\url{https://CRAN.R-project.org/package=DirStats}}. For computational simplification of simulations, the same bandwidth considered for computing $f_n$ is also used for estimate the corresponding $n_c$ density functions $f_{n,j}  $.

\begin{figure*}\vspace*{-.56cm}   
	\begin{picture}(-200,410)
 	\put(-90,195){\includegraphics[scale=.5]{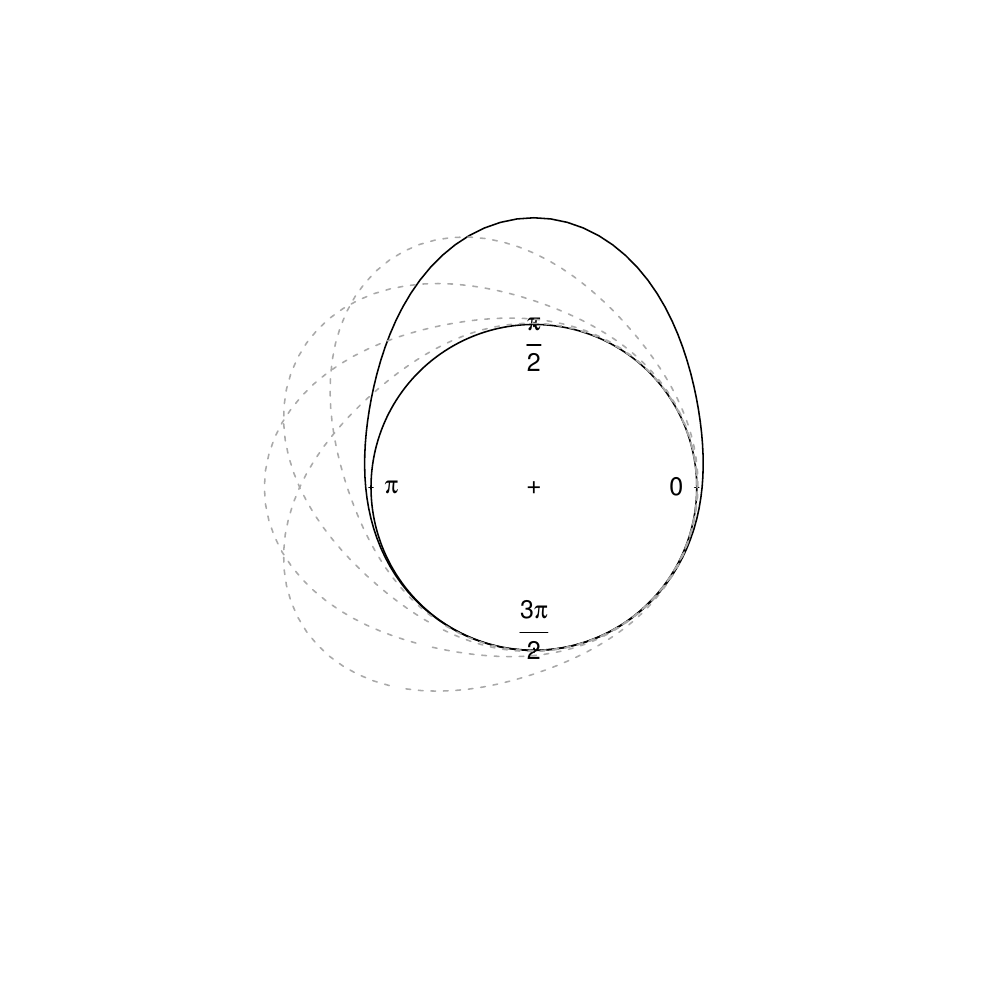}}
	\put(40,200){\includegraphics[scale=.5]{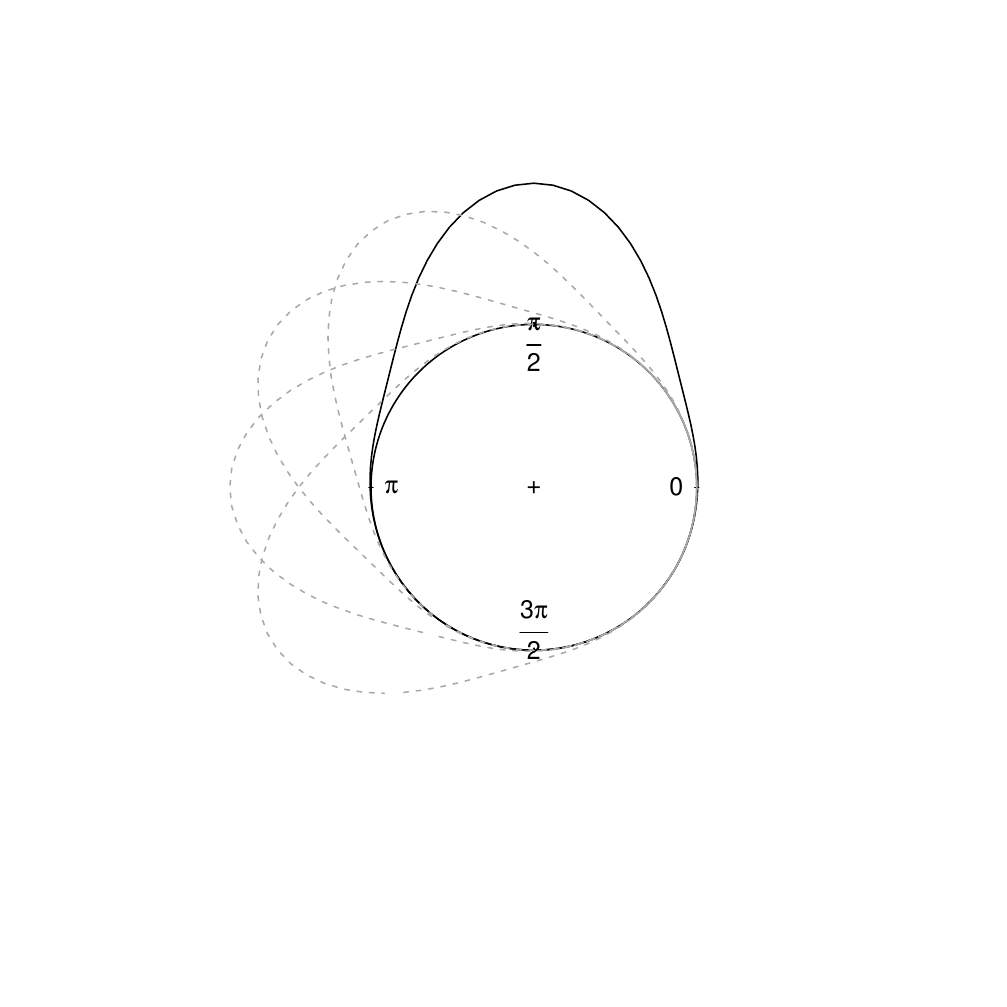}}
		\put(175,200){\includegraphics[scale=.5]{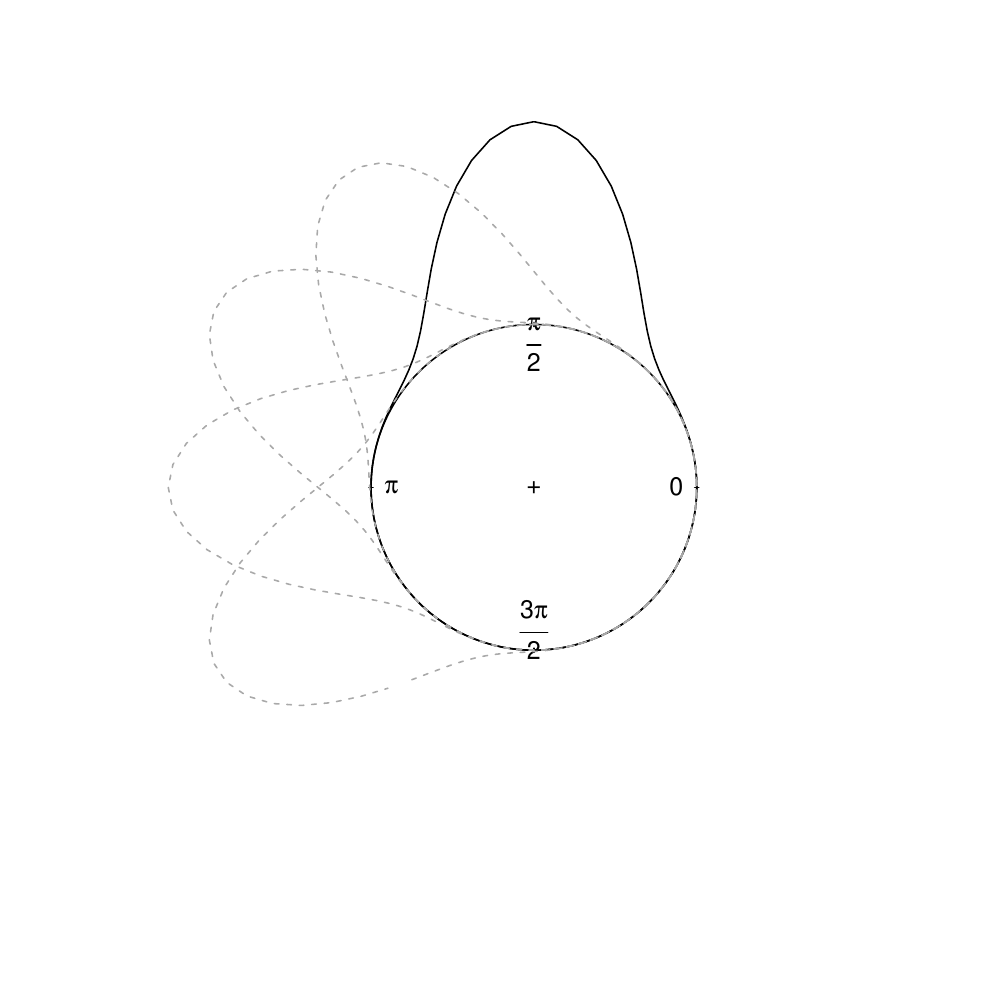}}
		 	\put(375,283){\includegraphics[scale=.33]{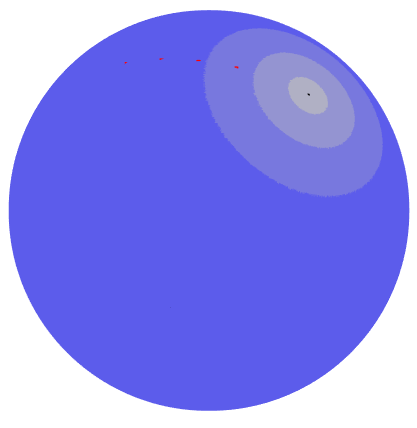}}
			\put(-90,65){\includegraphics[scale=.5]{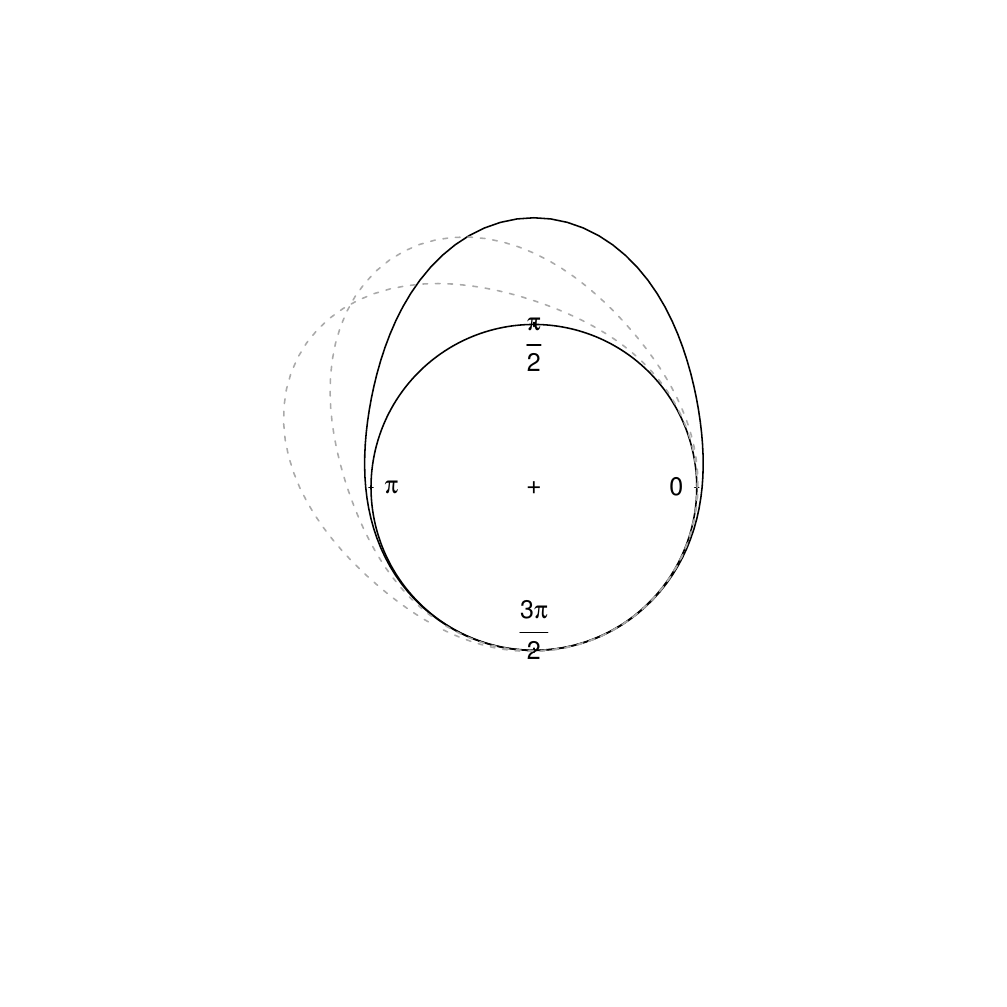}}
	\put(40,65){\includegraphics[scale=.5]{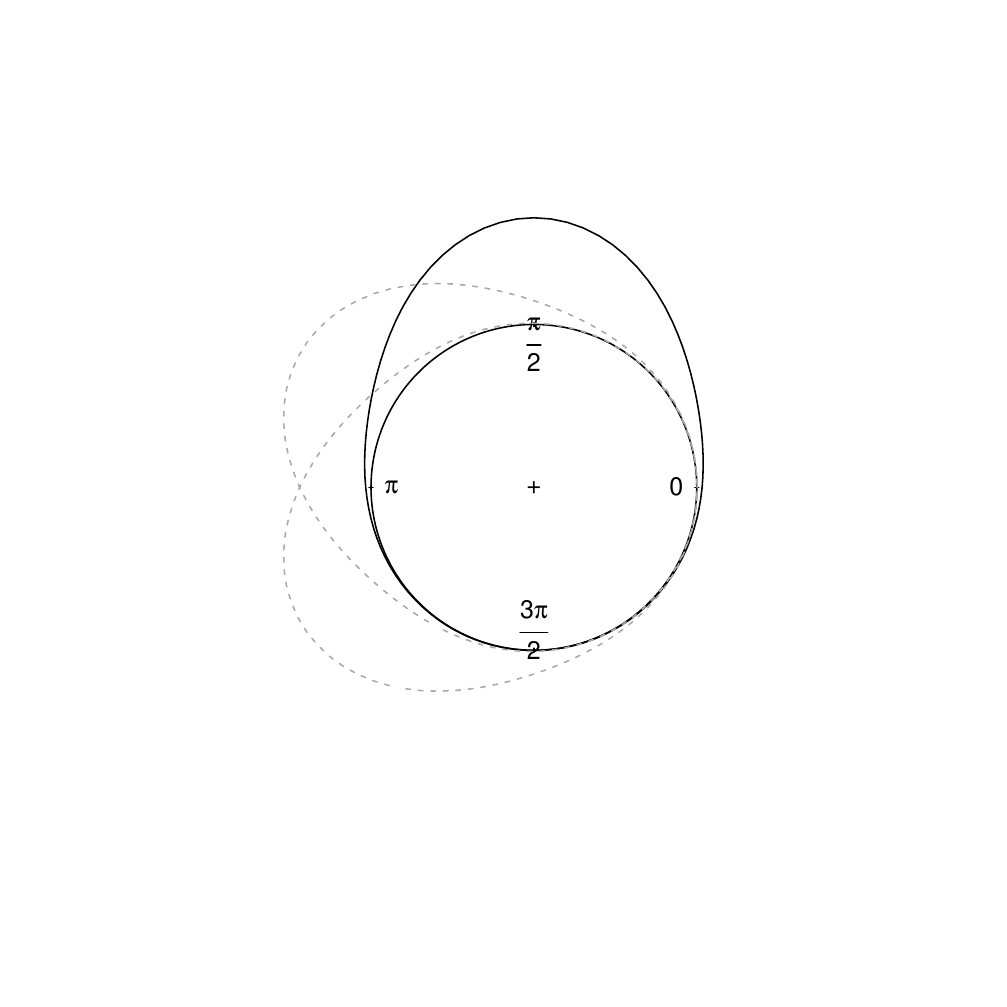}}
		\put(174,65){\includegraphics[scale=.5]{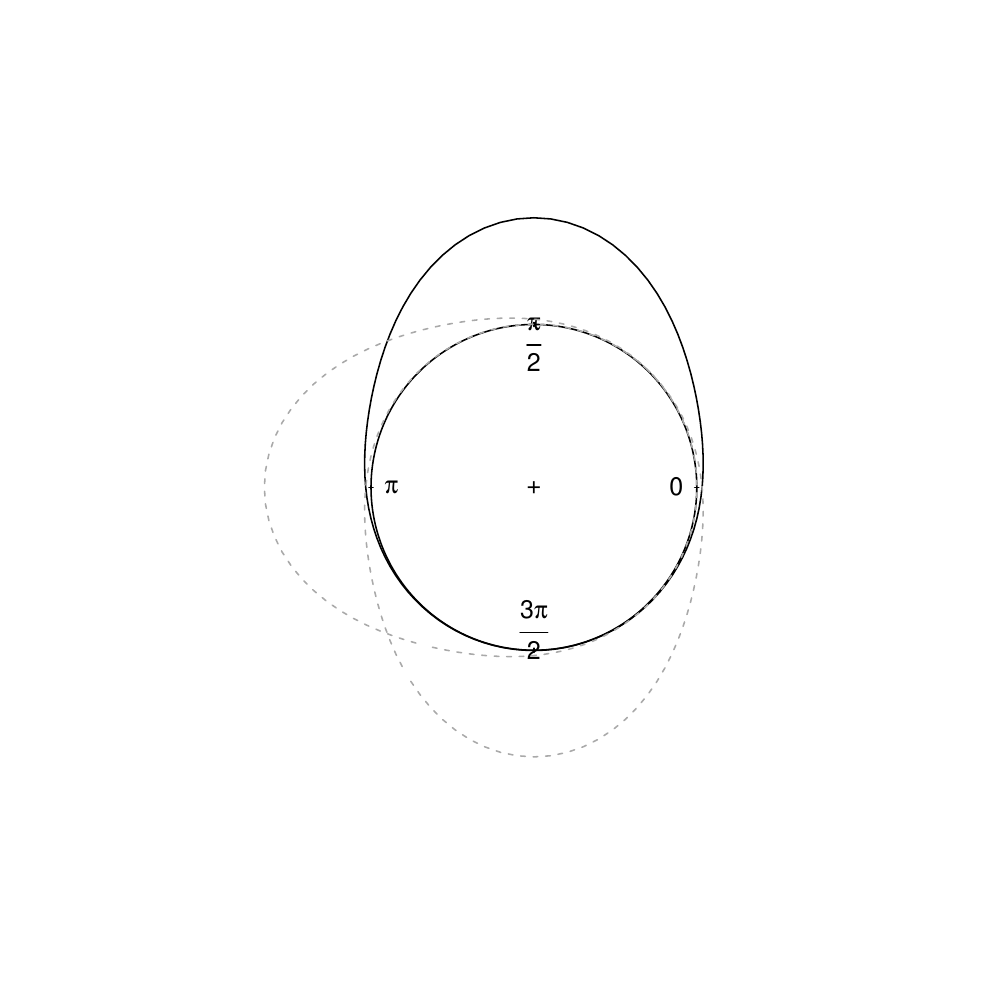}}
		\put(300,65){\includegraphics[scale=.5]{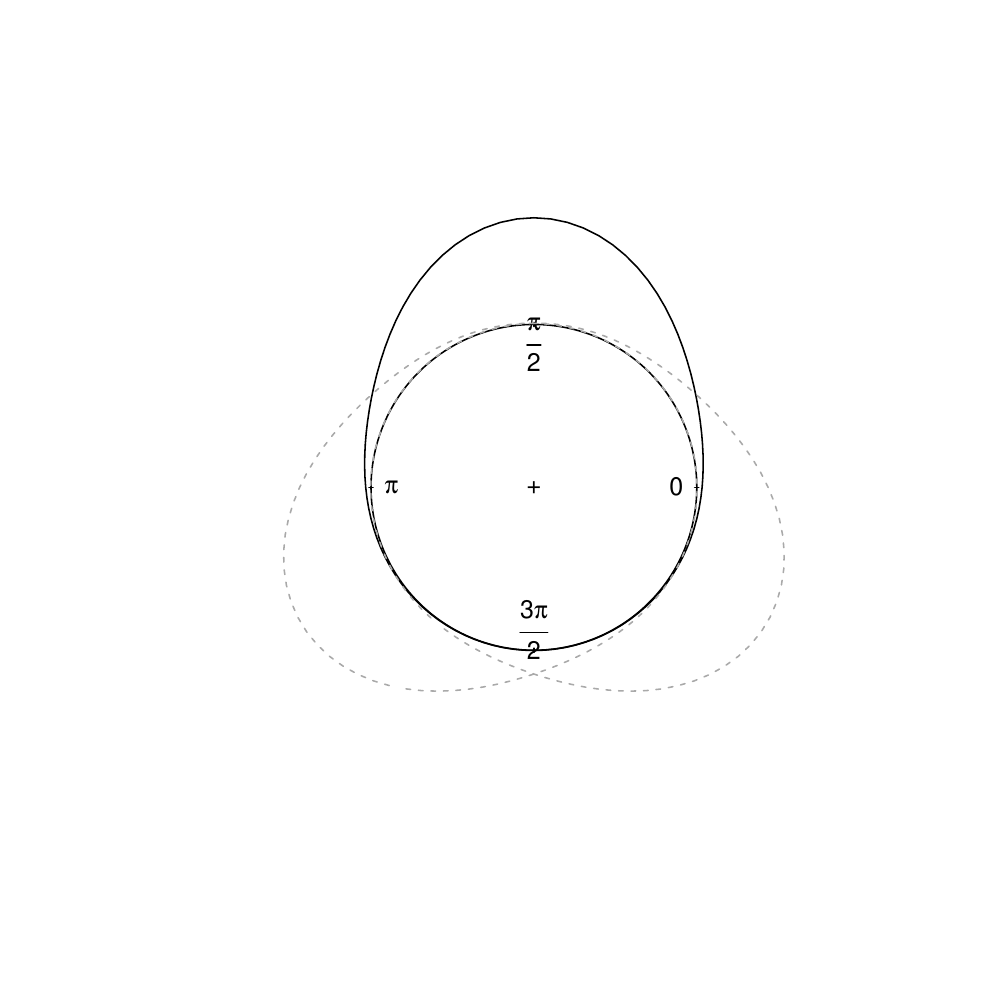}}
\end{picture}\vspace*{-5cm}\\  
     \caption{Circular and spherical von Mises-Fisher density models for simulations when the existence of two (first row) and three (second row) population groups is
assumed.}
     \label{modelossimus}
 \end{figure*}

The 8 simulation scenarios considered from circular and spherical von Mises-Fisher densities are shown in Figure \ref{modelossimus}. Specifically, first row of Figure \ref{modelossimus} contains the simulation models when the existence of two population groups is assumed. Columns from 1 to 3 (first row) show the circular models considered in this case for values of the concentration parameter $3$, $5$ and $10$, respectively. The black curve represents the generating density of first group for comparison with mean direction $\mu_1=\pi/2$. Gray densities correspond to the second successive groups with mean directions $\mu_2=\mu_1+\pi/6$, $\mu_2=\mu_1+2\pi/6$, $\mu_2=\mu_1+3\pi/6$ and $\mu_2=\mu_1+4\pi/6$, respectively. As for column 4 (first row), it shows the only spherical scenario considered in this work. The spherical model represented corresponds to a von Mises-Fisher density of the first group for comparison. Its concentration parameter is $20$ and mean direction $\mu_1=(45,90)=(\pi/4,\pi/2)$. Red points on this picture represent the mean directions $\mu_2$ of the density models corresponding to second groups. In this case, $\mu_2=(\pi/4+\pi/9,\pi/2)$, $\mu_2=(\pi/4+\pi/6,\pi/2)$, $\mu_2=(\pi/4+2\pi/9,\pi/2)$ and $\mu_2=(\pi/4+5\pi/18,\pi/2)$.
Second row of Figure \ref{modelossimus} contains the circular von Mises-Fisher models when the existence of three population groups is assumed. The concentration parameter is equal to 3 for all densities represented in the four columns. However, the mean directions are different: $\mu_1=\pi/2$, $\mu_2=\mu_1+\pi/6$ and $\mu_3=\mu_2+\pi/6$ (first column); $\mu_1=\pi/2$, $\mu_2=\mu_1+2\pi/6$ and $\mu_3=\mu_2+2\pi/6$ (second column); $\mu_1=\pi/2$, $\mu_2=\mu_1+\pi/2$ and $\mu_3=\mu_2+\pi/2$ (third column); and, finally,
$\mu_1=\pi/2$, $\mu_2=\mu_1+2\pi/3$ and $\mu_3=\mu_2+2\pi/3$ (fourth column).

Both in the spherical and circular scenarios, a total of $250$ simulations are performed. Specifically, a random sample of size $n$ is generated from each one of the (two or three) densities involved in each interaction. The considered values of $n$ are $750$, $1000$ and $1500$ in the circular setting and, $1000$ and $2000$ for spherical data. 
For each random sample, the classification method introduced in Section \ref{sec:classification} and $\kappa-$means are applied on sample constructed as the union of the (two or three) sets of observations involved. To assess and compare the performance of these classification methods, we evaluated the Adjusted Rand Index (ARI) proposed by \cite{hubert1985comparing} and used among others by \cite{stuetzle2003estimating} and \cite{azzalini2007clustering} for comparing competing non-directional clustering techniques.  

Tables \ref{tab:circular2gruposconcentracion3}, \ref{tab:circular2gruposconcentracion5} and \ref{tab:circular2gruposconcentracion10} show the means (M) and the standard deviations (SD) of the $250$ values of the ARI obtained when the existence of two groups is assumed in circular scenarios with concentration parameter equal to $3$, $5$ and $10$, respectively. The comparison of results for the same values of $\mu_2-\mu_1$ shows that the ARI is bigger as the concentration parameter increases because, as expected, classification becomes easier. Additionally, procedure proposed in Section \ref{sec:classification} is particularly less competitive than $2-$means when the concentration parameter is $3$ and $\mu_2-\mu_1$ takes the values $\pi/6$ or $2\pi/6$. However, results obtained show that our proposal is competitive when the concentration parameter increases and $\mu_2-\mu_1\geq 2\pi/6$.

 An objective comparison of two classification methodologies must take into account that the real value of $\kappa$ is given as an input here. Its value should be selected from data in practice. As for bandwidths, $h_1$ and $h_2$ (or even $h_6$) present a good performance when the concentration parameter is equal to $3$. If it takes the value $5$ or $10$, $h_1$ and $h_2$ continue to be the most competitive choice. However, all bandwidths considered in this study present a regular performance when the concentration parameter is $10$ as the means difference increases.

\begin{sidewaysfigure}
      \centering\vspace{-.8cm}
     \includegraphics[scale=0.3]{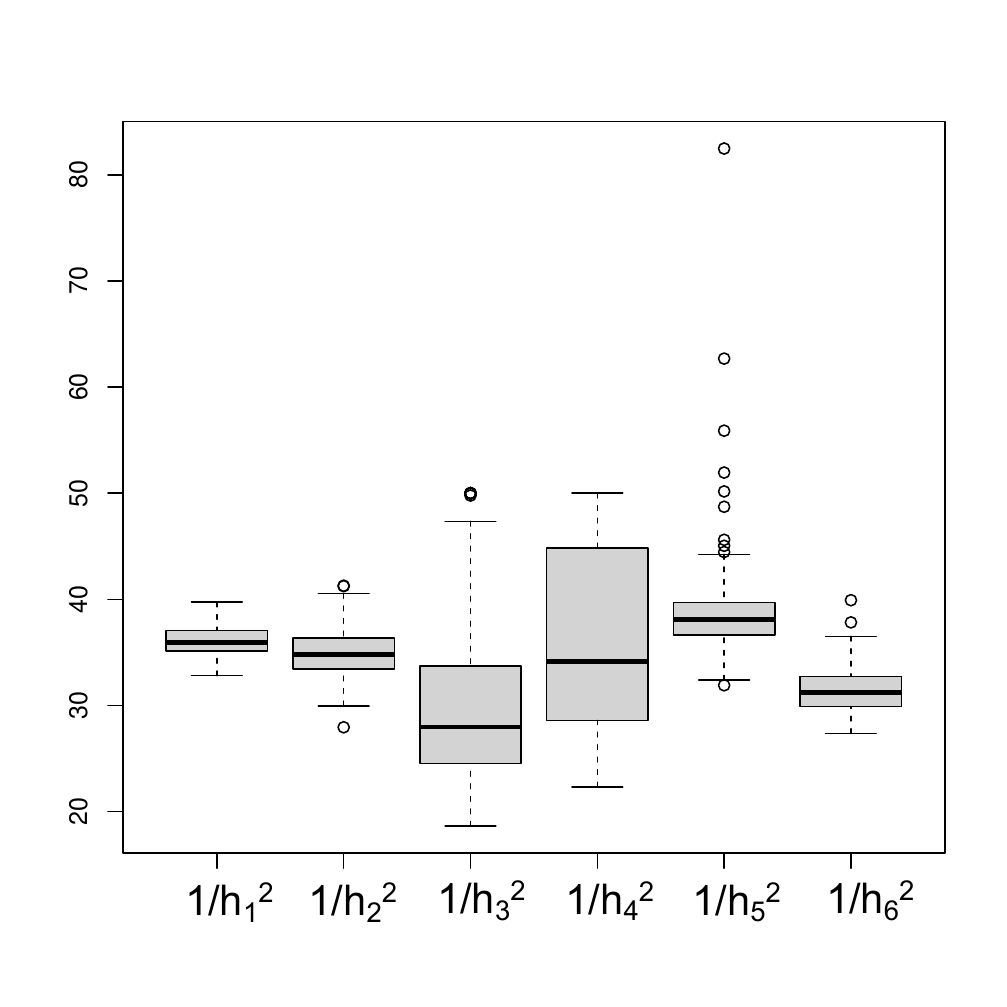} \includegraphics[scale=0.3]{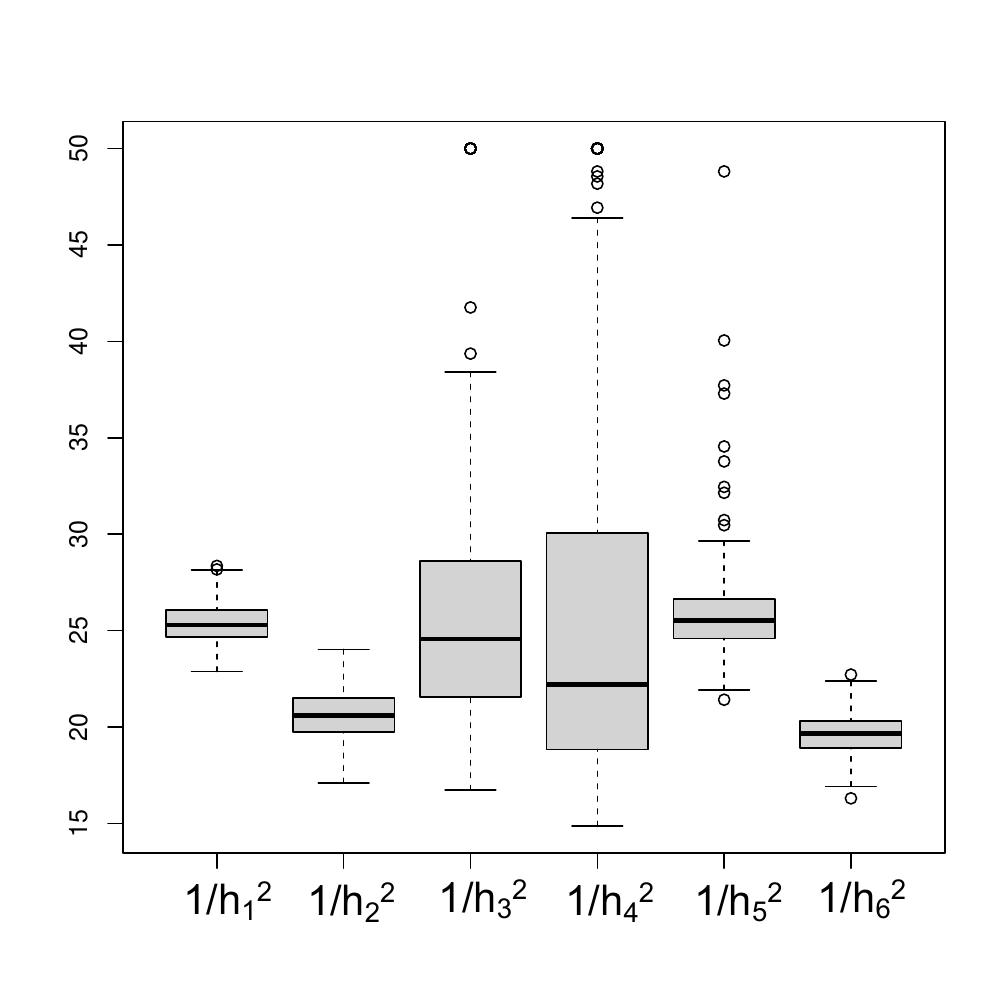}
      \includegraphics[scale=0.3]{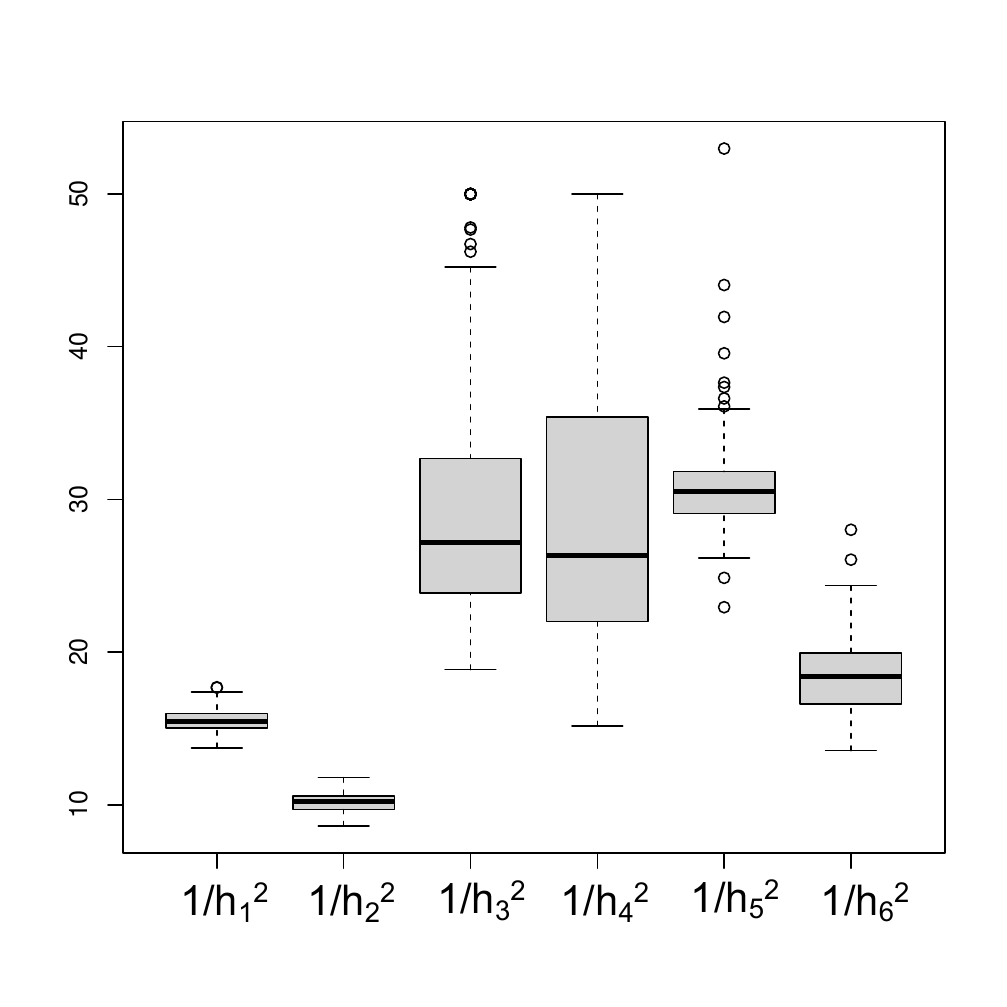}
       \includegraphics[scale=0.3]{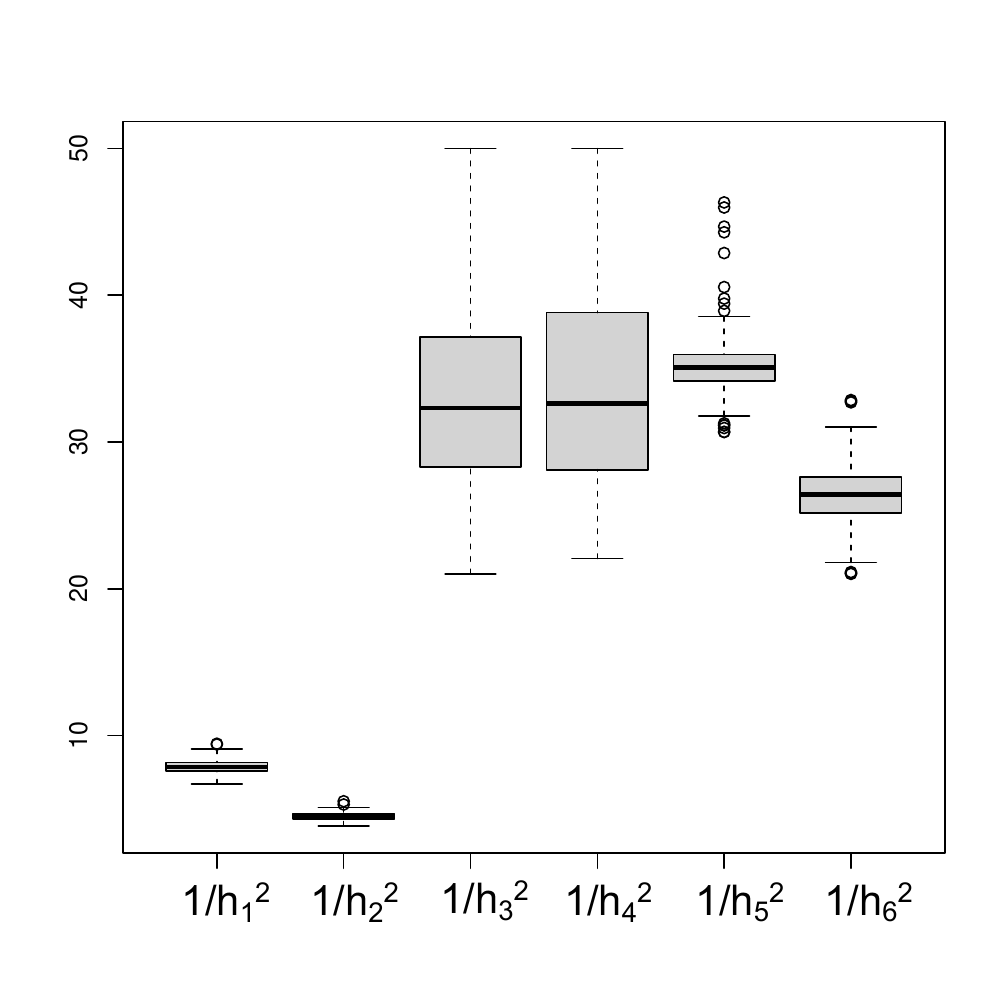}
     \caption{Estimated bandwidths from $h_1$ to $h_6$ for circular samples of size $750$ when the existence of two groups is assumed and the concentration parameter is equal to $3$ with $\mu_1-\mu_2=\pi/6$ (first column), $\mu_1-\mu_2=2\pi/6$ (second column), $\mu_1-\mu_2=3\pi/6$ (third column) and $\mu_1-\mu_2=2\pi/3$  (fourth column).}
     \label{fig:circular2gruposconcentracion3}
 \end{sidewaysfigure}

\begin{sidewaysfigure}
     \centering\vspace{-.8cm}
     \includegraphics[scale=0.3]{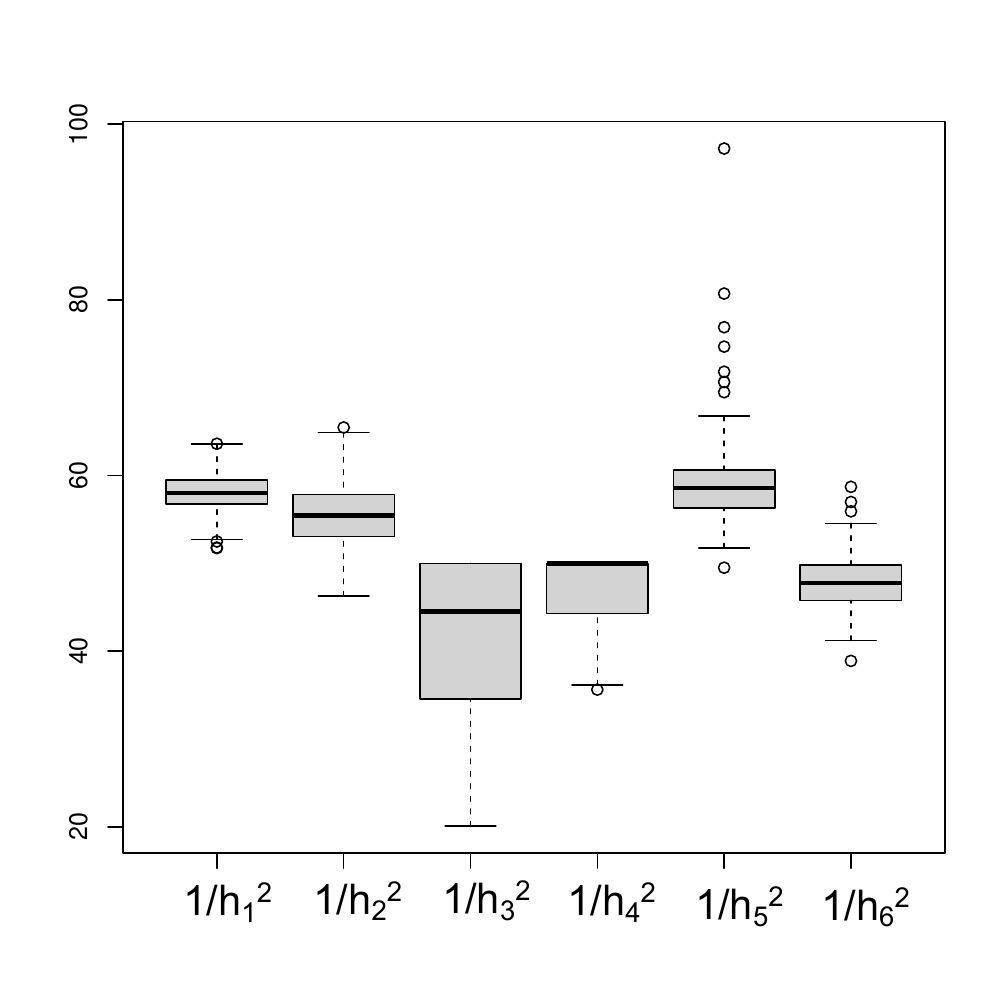} \includegraphics[scale=0.3]{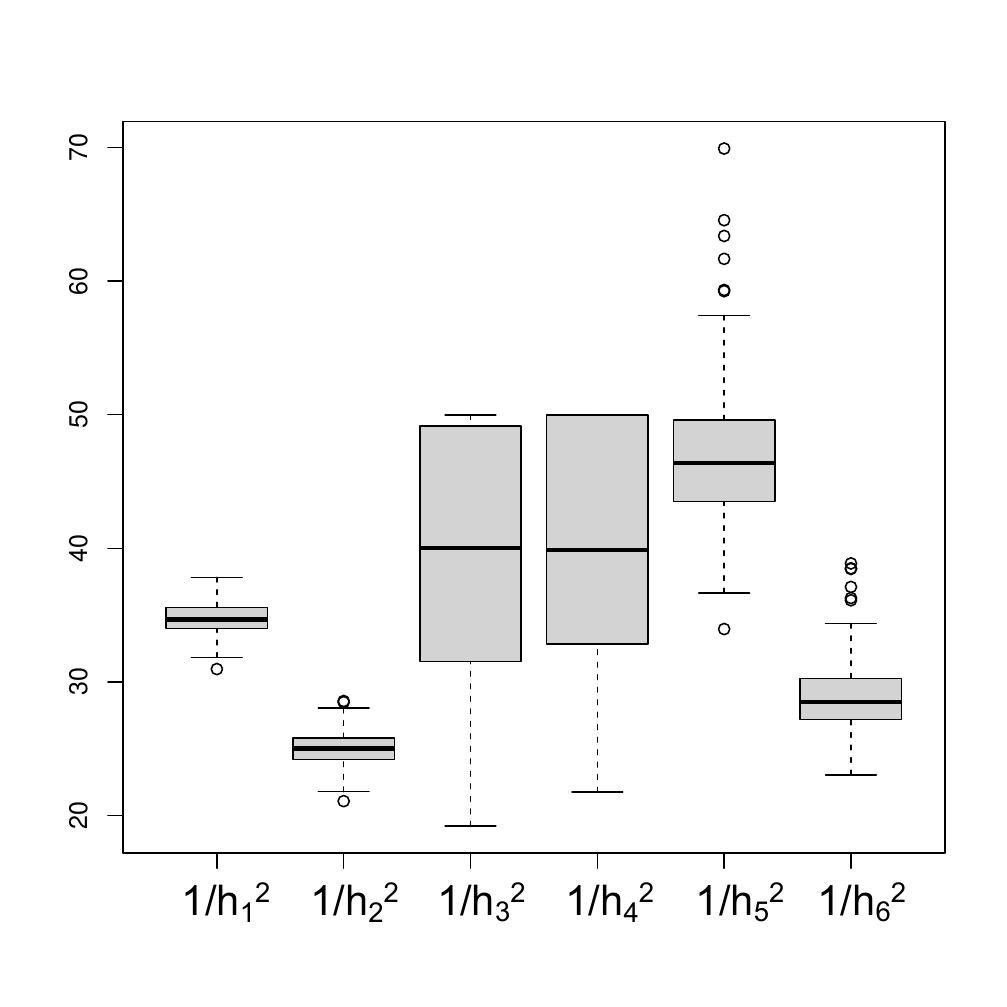}
      \includegraphics[scale=0.3]{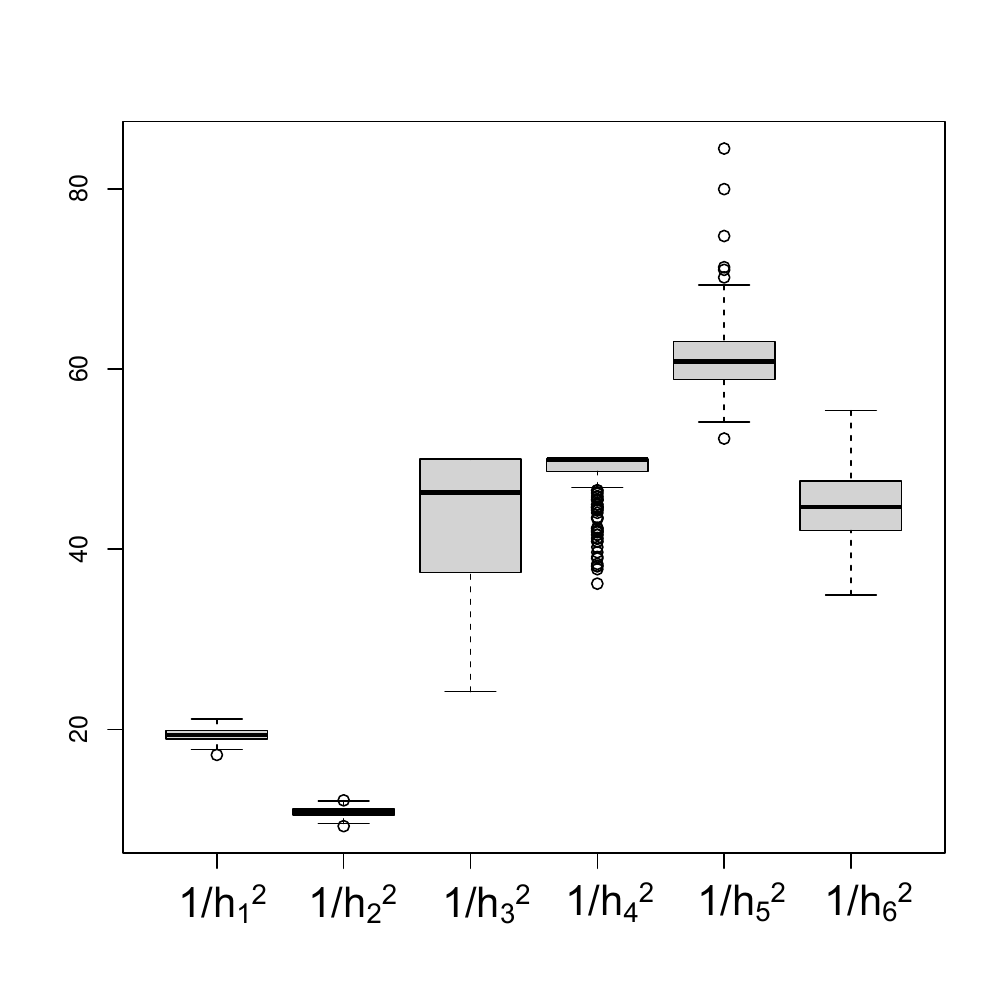}
       \includegraphics[scale=0.3]{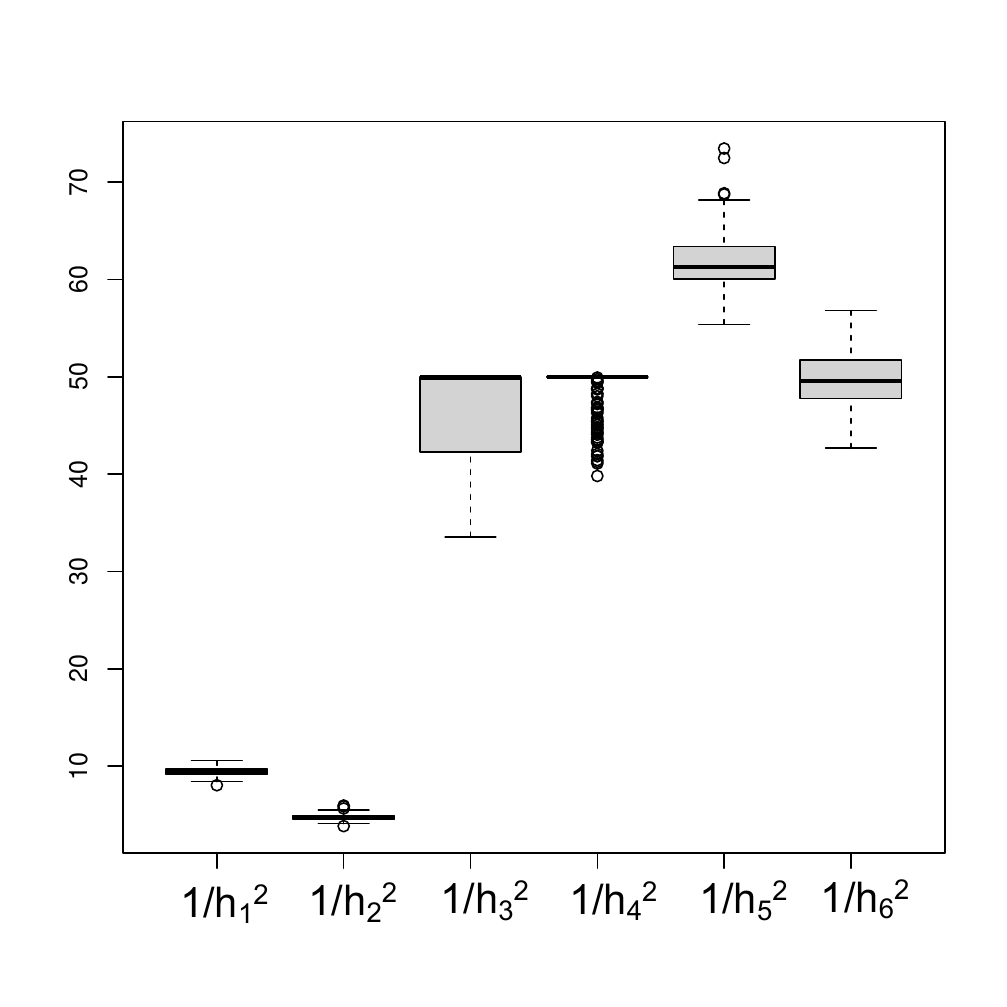}
     \caption{Estimated bandwidths from $h_1$ to $h_6$ for circular samples of size $750$ when the existence of two groups is assumed and the concentration parameter is equal to $5$ with $\mu_1-\mu_2=\pi/6$ (first column), $\mu_1-\mu_2=2\pi/6$ (second column), $\mu_1-\mu_2=3\pi/6$ (third column) and $\mu_1-\mu_2=2\pi/3$  (fourth column).}
     \label{fig:circular2gruposconcentracion5}
 \end{sidewaysfigure}

Additionally, Figures \ref{fig:circular2gruposconcentracion3} and \ref{fig:circular2gruposconcentracion5} contains the boxplots obtained from the values of the six estimated concentration parameters $1/h_i^2$ ($i=1,\cdots,6$) when $n=750$, respectively. Specifically, first column contains the scenario where $\mu_2-\mu_1=\pi/6$; second column, $\mu_2-\mu_1=2\pi/6$; third column, $\mu_2-\mu_1=3\pi/6$ and fourth column, $\mu_2-\mu_1=2\pi/3$. Graphical representations show that values of $h_1$ and $h_2$ 
 (the most competitive ones in these scenarios) are clearly bigger than the obtained for the rest of bandwidths as $\mu_2-\mu_1$ increases.

Table \ref{tab:circular3gruposconcentracion3} shows the means and the standard deviations of the $250$ values of the ARI obtained when the existence of three groups is assumed in circular scenarios with concentration parameter equal to $3$. In this case, comparison between the ARIs obtained from the density-based classification algorithm and $3-$means shows that our proposal is competitive when $h_3$, $h_4$, $h_5$ and $h_6$ are used as bandwidths selectors specially when means differences are bigger than $2\pi/6$.

Table \ref{tab:esferico2grupos} contains the means and the standard deviations of the $250$ values of the ARI obtained when the existence of two groups is assumed in spherical scenarios. If $\mu_2-\mu_1=(\pi/6, 0)$, $h_8$ and $h_9$ (or even $h_7$) present a competitive performance. However, $h_3$ (jointly with $h_8$) provides the highest values of ARI as the first component of vector $\mu_2-\mu_1$ increases. 

\begin{sidewaystable}
\sidewaystablefn%
\begin{center}
\begin{minipage}{\textheight}
     \small{
     \begin{tabular}{cccccccccccccccc}
       \hline
    & $n$&\multicolumn{2}{c}{$h_1$}& \multicolumn{2}{c}{$h_2$}&\multicolumn{2}{c}{$h_3$}&\multicolumn{2}{c}{$h_4$}&\multicolumn{2}{c}{$h_5$}&\multicolumn{2}{c}{$h_6$}&\multicolumn{2}{c}{$2-$means}\\
    $\mu_2-\mu_1$& &M&SD&M&SD&M&SD&M&SD&M&SD&M&SD&M&SD\\
    \hline      $\pi/6$&750&0.001&0.003&0.001&0.003&0.001&0.004&0.001&0.007&0.001&0.003&0.000&0.002&0.108&0.017\\
$\pi/6$&1000&0.001&0.002&0.001&0.002&0.001&0.006&0.001&0.007&0.001&0.007&0.000&0.002&0.109&0.014\\
$\pi/6$&1500&0.001&0.003&0.001&0.003&0.001&0.002&0.001&0.002&0.001&0.003&0.001&0.003&0.109&0.012\\
\hline
$2\pi/6$&750&0.034&0.105&0.027&0.096&0.031&0.100&0.028&0.097&0.040&0.111&0.023&0.089&0.358&0.023\\
$2\pi/6$&1000&0.043&0.116&0.026&0.092&0.041&0.113&0.034&0.106&0.039&0.111&0.022&0.087&0.360&0.021\\
$2\pi/6$&1500&0.048&0.121&0.027&0.094&0.037&0.107&0.041&0.116&0.047&0.120&0.030&0.099&0.361&0.018\\
\hline
$3\pi/6$&750&0.588&0.091&0.586&0.090&0.498&0.253&0.501&0.253&0.478&0.263&0.566&0.142&0.612&0.025\\
$3\pi/6$&1000&0.588&0.097&0.600&0.049&0.489&0.258&0.501&0.248&0.468&0.273&0.557&0.168&0.614&0.021\\
$3\pi/6$&1500&0.594&0.101&0.610&0.020&0.492&0.265&0.514&0.247&0.456&0.291&0.550&0.187&0.615&0.018\\
\hline
$2\pi/3$&750&0.787&0.021&0.787&0.021&0.621&0.321&0.655&0.293&0.624&0.318&0.715&0.224&0.789&0.021\\
$2\pi/3$&1000&0.788&0.018&0.788&0.018&0.643&0.304&0.657&0.294&0.632&0.311&0.709&0.233&0.789&0.018\\
$2\pi/3$&1500&0.790&0.014&0.790&0.014&0.682&0.265&0.708&0.235&0.659&0.287&0.726&0.215&0.791&0.014\\
  \hline
     \end{tabular}
     }
     \caption{Means (M) and the standard deviations (SD) of the $250$ values of the ARI computed in circular simulation scenarios when the existence of two groups is assumed and the concentration parameter is equal to $3$.}
     \label{tab:circular2gruposconcentracion3}
\end{minipage}
     \end{center}
 \end{sidewaystable}
 
\begin{sidewaystable}
\sidewaystablefn%
\begin{center}
\begin{minipage}{\textheight}
     \small{
     \begin{tabular}{cccccccccccccccc}
       \hline
    & $n$&\multicolumn{2}{c}{$h_1$}& \multicolumn{2}{c}{$h_2$}&\multicolumn{2}{c}{$h_3$}&\multicolumn{2}{c}{$h_4$}&\multicolumn{2}{c}{$h_5$}&\multicolumn{2}{c}{$h_6$}&\multicolumn{2}{c}{$2-$means}\\
    $\mu_2-\mu_1$& $n$&M&SD&M&SD&M&SD&M&SD&M&SD&M&SD&M&SD\\
    \hline $\pi/6$&750&0.005&0.030&0.003&0.023&0.002&0.018&0.002&0.018&0.006&0.033&0.001&0.012&0.184&0.021\\
$\pi/6$&1000&0.006&0.033&0.004&0.027&0.000&0.000&0.001&0.011&0.006&0.033&0.001&0.011&0.184&0.018\\
$\pi/6$&1500&0.004&0.026&0.003&0.022&0.000&0.000&0.000&0.000&0.004&0.025&0.001&0.011&0.184&0.016\\ \hline
$2\pi/6$&750&0.331&0.252&0.292&0.261&0.321&0.255&0.323&0.262&0.349&0.254&0.303&0.259&0.543&0.024\\
$2\pi/6$&1000&0.345&0.242&0.305&0.258&0.342&0.250&0.343&0.255&0.373&0.254&0.326&0.251&0.544&0.021\\
$2\pi/6$&1500&0.340&0.253&0.332&0.255&0.325&0.258&0.342&0.253&0.332&0.268&0.337&0.255&0.545&0.018\\ \hline
$3\pi/6$&750&0.757&0.206&0.806&0.075&0.577&0.368&0.569&0.371&0.552&0.374&0.598&0.356&0.814&0.021\\
$3\pi/6$&1000&0.722&0.257&0.806&0.075&0.607&0.354&0.604&0.355&0.478&0.398&0.584&0.365&0.814&0.018\\
$3\pi/6$&1500&0.688&0.297&0.815&0.015&0.563&0.377&0.573&0.373&0.447&0.403&0.509&0.393&0.815&0.015\\ \hline
$2\pi/3$&750&0.938&0.012&0.939&0.012&0.647&0.433&0.651&0.432&0.593&0.451&0.640&0.436&0.939&0.012\\
$2\pi/3$&1000&0.938&0.011&0.938&0.011&0.682&0.418&0.664&0.427&0.535&0.464&0.644&0.435&0.938&0.011\\
$2\pi/3$&1500&0.939&0.009&0.939&0.009&0.688&0.417&0.688&0.417&0.489&0.466&0.570&0.457&0.939&0.009\\
  \hline
     \end{tabular}}
     \caption{Means (M) and the standard deviations (SD) of the $250$ values of the ARI computed in circular simulation scenarios when the existence of two groups is assumed and the concentration parameter is equal to $5$.}
     \label{tab:circular2gruposconcentracion5}
\end{minipage}
     \end{center}
 \end{sidewaystable}
  
\begin{sidewaystable} 
\sidewaystablefn%
\begin{center}
\begin{minipage}{\textheight}
     \small{
\begin{tabular}{cccccccccccccccc}
       \hline
      & &\multicolumn{2}{c}{$h_1$}& \multicolumn{2}{c}{$h_2$}&\multicolumn{2}{c}{$h_3$}&\multicolumn{2}{c}{$h_4$}&\multicolumn{2}{c}{$h_5$}&\multicolumn{2}{c}{$h_6$}&\multicolumn{2}{c}{$2-$means}\\
   $\mu_2-\mu_1$&$n$ &M&SD&M&SD&M&SD&M&SD&M&SD&M&SD&M&SD\\
    \hline
 $ $&750&0.026&0.092&0.022&0.086&0.003&0.031&0.003&0.031&0.027&0.094&0.015&0.071&0.341&0.023\\
$\pi/6$&1000 &0.032&0.100&0.022&0.085&0.000&0.000&0.000&0.000&0.029&0.095&0.011&0.061&0.341&0.022\\
$ $ &1500&0.033&0.101&0.028&0.093&0.000&0.000&0.000&0.000&0.016&0.073&0.015&0.070&0.341&0.019\\\hline
$ $ &750&0.779&0.114&0.796&0.022&0.779&0.114&0.779&0.114&0.702&0.256&0.721&0.231&0.797&0.021\\
$ 2\pi/6$ &1000&0.771&0.142&0.787&0.089&0.765&0.158&0.765&0.158&0.711&0.244&0.718&0.236&0.798&0.019\\
$ $ &1500&0.760&0.172&0.789&0.089&0.770&0.150&0.770&0.150&0.705&0.256&0.705&0.256&0.800&0.015\\\hline
$ $ &750&0.958&0.087&0.966&0.009&0.950&0.122&0.950&0.122&0.858&0.305&0.866&0.296&0.966&0.009\\
$3\pi/6$ &1000&0.962&0.062&0.966&0.008&0.927&0.190&0.927&0.190&0.854&0.310&0.854&0.310&0.966&0.008\\
$ $ &1500&0.962&0.062&0.962&0.062&0.931&0.180&0.931&0.180&0.865&0.295&0.865&0.295&0.966&0.007\\\hline
$ $ &750&0.996&0.003&0.996&0.003&0.980&0.125&0.980&0.125&0.908&0.283&0.908&0.283&0.996&0.003\\
$2\pi/3$ &1000&0.996&0.003&0.996&0.003&0.956&0.196&0.956&0.196&0.888&0.310&0.888&0.310&0.996&0.003\\
$ $ &1500&0.993&0.063&0.996&0.002&0.961&0.186&0.961&0.186&0.881&0.320&0.881&0.320&0.997&0.002\\
  \hline
     \end{tabular}}
     
     \caption{Means (M) and the standard deviations (SD) of the $250$ values of the ARI computed in circular simulation scenarios when the existence of two groups is assumed and the concentration parameter is equal to $10$.}\label{tab:circular2gruposconcentracion10}\end{minipage}
     \end{center}
 \end{sidewaystable}

\begin{sidewaystable}
\sidewaystablefn%
\begin{center}
\begin{minipage}{\textheight}
     \small{\begin{tabular}{ccccccccccccccc}
       \hline
    $\mu_{2}-\mu_1$&\multicolumn{2}{c}{$h_1$}& \multicolumn{2}{c}{$h_2$}&\multicolumn{2}{c}{$h_3$}&\multicolumn{2}{c}{$h_4$}&\multicolumn{2}{c}{$h_5$}&\multicolumn{2}{c}{$h_6$}&\multicolumn{2}{c}{$3-$means}\\
   $\mu_{3}-\mu_2$  &M&SD&M&SD&M&SD&M&SD&M&SD&M&SD&M&SD\\
    \hline $\pi/6$&0.003&0.023&0.001&0.014&0.003&0.021&0.003&0.023&0.002&0.018&0.001&0.015&0.122&0.010\\
$2\pi/6$&0.114&0.159&0.044&0.116&0.197&0.162&0.149&0.166&0.180&0.162&0.100&0.154&0.371&0.016\\
$3\pi/6$&0.274&0.183&0.030&0.103&0.592&0.051&0.594&0.054&0.592&0.051&0.587&0.054&0.605&0.017\\
$2\pi/3$&0.000&0.000&0.259&0.197&0.708&0.033&0.707&0.032&0.707&0.025&0.703&0.048&0.709&0.016\\
  \hline\end{tabular}}
     \caption{Means (M) and the standard deviations (SD) of the $250$ values of the ARI computed in circular simulation scenarios when the existence of three groups is assumed, the concentration parameter is equal to $3$ and $n=750$.}
 \label{tab:circular3gruposconcentracion3} \end{minipage}
     \end{center}
 \end{sidewaystable}
 
\begin{sidewaystable}
\sidewaystablefn%
\begin{center}
\begin{minipage}{\textheight}
     \small{
\begin{tabular}{cccccccccccccc}
       \hline
    &    &\multicolumn{2}{c}{$h_3$}&  \multicolumn{2}{c}{$h_4$}&\multicolumn{2}{c}{$h_7$}&\multicolumn{2}{c}{$h_8$}&\multicolumn{2}{c}{$h_9$}&\multicolumn{2}{c}{$2-$means}\\
  $\mu_2-\mu_1$  &$n$ &M&SD&M&SD&M&SD&M&SD&M&SD&M&SD\\
    \hline
$(\pi/9,0)$&1000&0.000&0.000&0.034&0.094&0.030&0.088&0.014&0.062&0.034&0.093&0.307&0.019\\
&2000&0.000&0.000&0.019&0.070&0.021&0.075&0.012&0.059&0.016&0.066&0.310&0.014\\
\hline
$(\pi/6,0)$&1000&0.175&0.259&0.392&0.243&0.397&0.235&0.408&0.226&0.384&0.251&0.558&0.021\\
&2000&0.281&0.278&0.390&0.261&0.407&0.247&0.407&0.234&0.406&0.255&0.560&0.016\\
\hline
$(2\pi/9,0)$&1000&0.759&0.019&0.585&0.323&0.631&0.283&0.705&0.193&0.640&0.280&0.761&0.018\\
&2000&0.759&0.014&0.488&0.361&0.534&0.344&0.630&0.281&0.532&0.345&0.761&0.013\\
\hline
$(5\pi/18,0)$&1000&0.888&0.014&0.671&0.368&0.710&0.347&0.801&0.259&0.698&0.349&0.889&0.015\\
&2000&0.887&0.010&0.522&0.421&0.566&0.413&0.718&0.344&0.575&0.407&0.888&0.010\\
 \hline
     \end{tabular}    }
     \caption{Means (M) and the standard deviations (SD) of the $250$ values of the ARI computed in spherical simulation scenarios when the existence of two groups is assumed.}
 \label{tab:esferico2grupos}
 \end{minipage}
     \end{center}
 \end{sidewaystable}
 
 \section{Exoplanets clustering analysis}\label{realdata}

As of 1 October 2022, there exist 5,197 confirmed extrasolar planets
in a total of 3833 planetary systems (with 840 systems being composed by more than one exoplanet).
These discoveries have opened a recent page in the astronomy. Planets of the Solar
System can only be observed in their current state; however, the observation of
other planetary systems through the years could reveal details on their formation
and evolution. Following \cite{hung2015intuitive}, unsupervised clustering techniques
for directional data are a powerful exploratory tool for grouping exoplanets data
and to showing hidden structural information.
 
An analogue clustering analysis to the presented in \cite{hung2015intuitive} for exoplanets is performed here from density-based clustering techniques introduced in this work. Exoplanets dataset has been downloaded from the website of \emph{The Extrasolar
Planets Encyclopaedia}\footnote{\url{http://exoplanet.eu/catalog/} - Download date: April 11th, 2022}.  For each exoplanet registered, this dataset contains several relevant astronomical variables such as projected mass ($M_p$), orbital period ($P$), semimature axis ($a$), orbital
eccentricity ($e$), stellar metallicity ($[Fe/H]$) and stellar mass ($M_s$).

 \begin{figure*}
     \centering
     \includegraphics[scale=0.28]{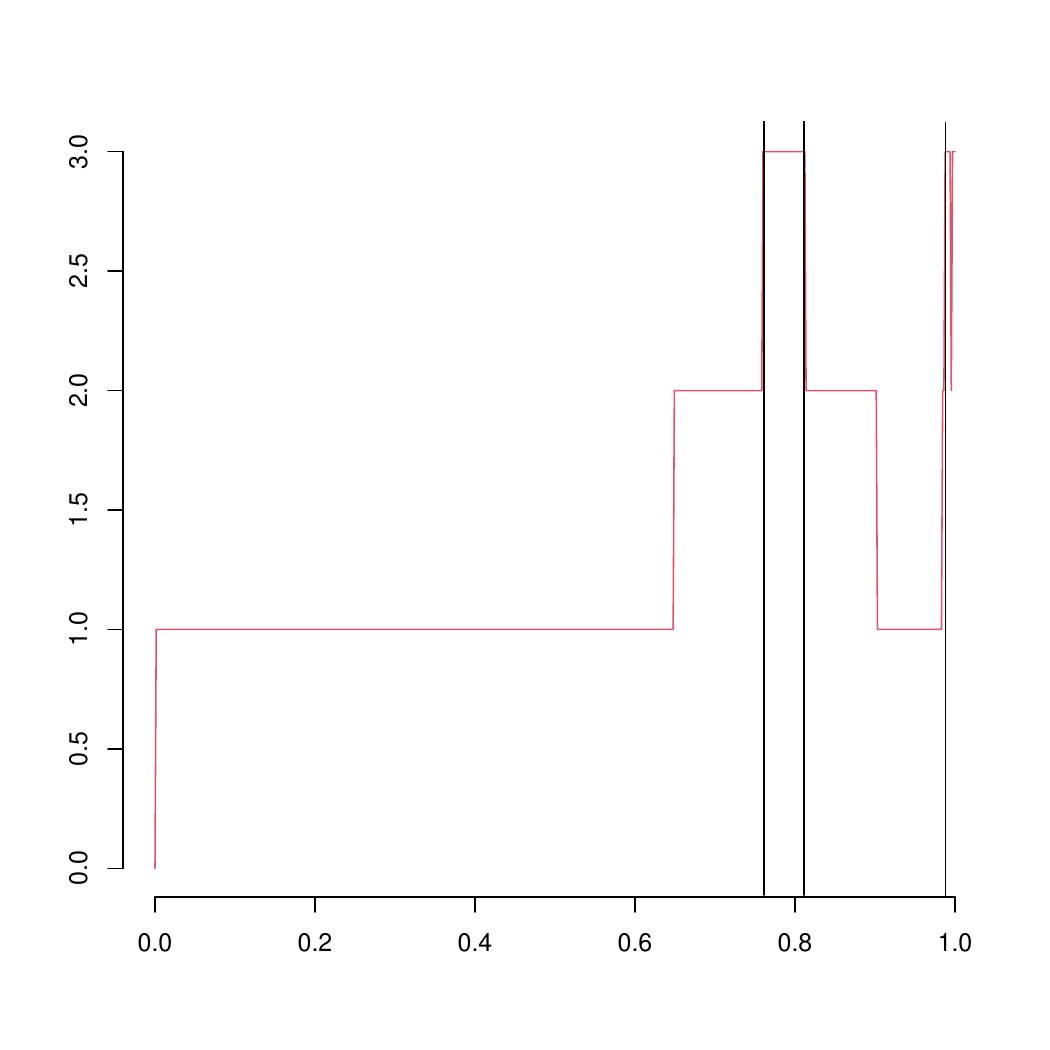}\includegraphics[scale=0.28]{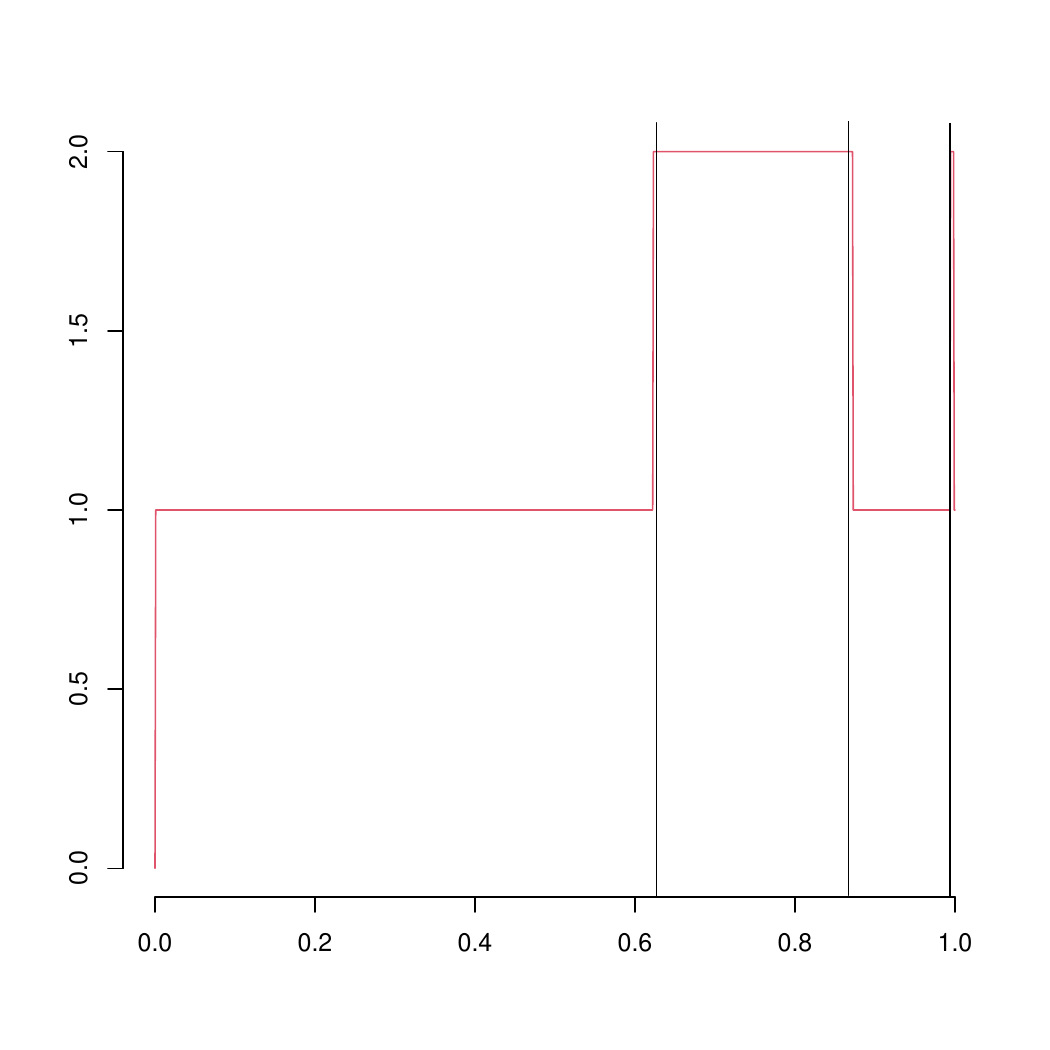}\includegraphics[scale=0.28]{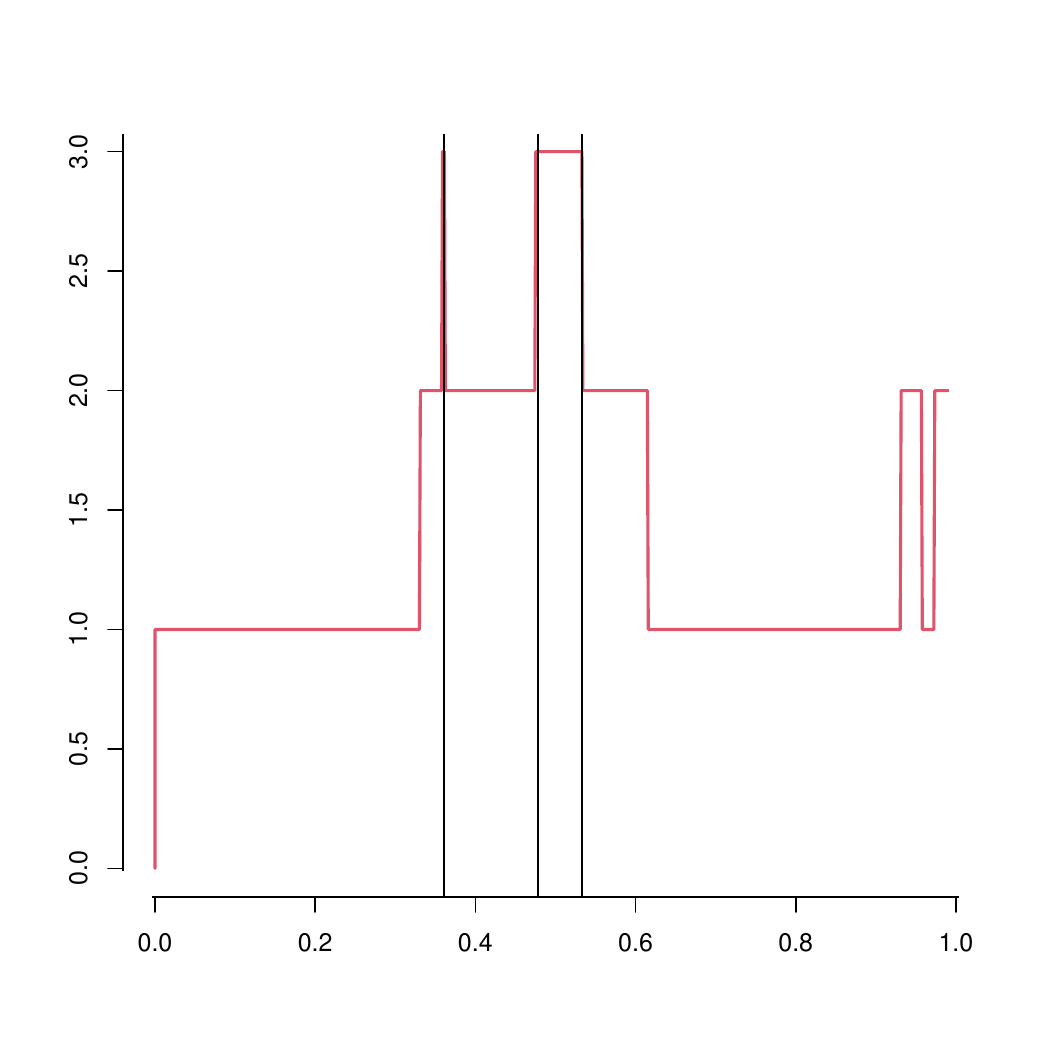}
     \caption{Empirical mode functions obtained from kernel density estimators in $S^1$ for exoplanets discovered in 2014 or before (left) and 2021 or before (center) by using a cross validation bandwidths. Empirical mode functions obtained from kernel density estimators in $S^4$ for exoplanets discovered in 2021 by selecting the bandwidth through a rule-of-thumb approach (right).}
     \label{fig:exo}
 \end{figure*} 

Several works in astronomical literature point out the existence of correlation between the variables $P$ and $M_p$ (for instance, see \citealp{ref1}, \citealp{ref2} and \citealp{ref3}). Following the strategy in \cite{hung2015intuitive}, we will check where exoplanets groups on these two features
are located. Density-based algorithm introduced in Section \ref{sec:classification} will be applied to the data $(\ln{M_p}, \ln{P})$ on $S^1$. Specifically, two different analysis in $S^1$ will be performed. Mainly, for comparison to results in \cite{hung2015intuitive}, a total of $648$ complete observations corresponding to exoplanets discovered in 2014 or before will be used; then, the same analysis is repeated for the $1093$ exoplanets (with complete information) discovered in 2021 or before. Besides, \cite{marchi2007extrasolar} studied the existence of exoplanets clusters where the  correlation among variables $M_p$, $a$, $e$, $[Fe/H]$ and $M_s$ was considerably strong. Therefore, our clustering proposal and $\kappa-$means method will be also applied for the dataset $(M_p, a, e, [Fe/H],M_s)$ on $S^4$ in order to check the existence of such correlation.

Figure \ref{fig:exo} contains the empirical mode functions obtained from kernel density estimators (with cross validation bandwidths) in $S^1$ for exoplanets discovered in 2014 or before (left) and in 2021 or before (center). Black vertical lines correspond to the different values of $1-\tau$ (and, therefore, of the threshold $\hat{f}_{\tau}$) that will be considered in this section for establishing the cluster cores. Remark that our clustering proposal identifies a maximum of three groups in 2014 and, two groups in 2021. Then, the number of clusters has decreased over time by showing a higher degree of grouping among them. Although four clusters were initially detected in 2014 by \cite{hung2015intuitive}, one of them was not representative because it contained an only exoplanet.

\begin{table*}
 \centering
 \small{
\begin{tabular}{ccccccc} 
 \hline
 Year & $1-\tau$&Cluster core & Center $(\ln M_p ,\ln P)$ & Members & Correlation & \textit{p-value}  \\
 \hline
 \multirow{9}*{2014} & \multirow{3}*{0.76}& C1 & $(0.724, 6.327)$ & 427 & 0.143 &  $<0.01$\\
                 &    & C2 & $( -1.386,  3.390)$ & 7 & -0.652 & 0.112 \\
                 &    & C3 & $(-3.185 , 2.852)$ & 58 & -0.505  &$<0.01$ \\
 \cline{2-7}
 \multirow{3}*{ }& \multirow{3}*{0.81}& C1 & $(0.704 ,6.292)$     & 432 & 0.146 & $<0.01$ \\
                 &   & C2 & $(-1.547,  3.865)$ & 29 & -0.693  & $<0.01$\\
                 &   & C3 & $(-3.211 , 2.860)$ & 65 & -0.431 & $<0.01$ \\
   
 \cline{2-7}
 \multirow{3}*{ }& \multirow{3}*{0.99}& C1 & $(2.681, 0.756 )$      & 3  &  0.913 & 0.268 \\
              &      & C2 & $( 1.660 ,1.265 )$  & 4  &0.992    & $<0.01$\\
               &     & C3 & $(-0.344 , 5.219 )$ &637 &0.175 & $<0.01$ \\
  \hline
 
 \multirow{6}*{2021} & \multirow{2}*{0.62}& C1 & $(0.786 ,  6.273)$& 671 & 0.220 & $<0.01$ \\
                 &    & C2 & $( -3.378,  2.870)$ & 10 & -0.909 & $<0.01$ \\
 \cline{2-7}
  \multirow{2}*{ } & \multirow{2}*{0.87}& C1    & $(0.656 ,6.121)$& 717     & 0.224& $<0.01$ \\
                 &     & C2 & $( -3.081,  3.156)$ & 236 & -0.073  & 0.261 \\
 \cline{2-7}
\multirow{2}*{ } & \multirow{2}*{0.99}& C1 & $( 3.025 ,-2.014)$& 3 & 0.951 &  0.199 \\
                &    & C2 & $( -0.483,  4.973)$ & 1086 &  0.149 & $<0.01$\\
 \hline

\end{tabular}
 }
 \caption{Classification results in $S^1$ for exoplanets discovered up to 2014 and up to 2021.}
 \label{tab:exos1}
\end{table*}

\begin{figure*}
    \begin{picture}(-200,400)
        \put(-10,205){\includegraphics[scale=.4]{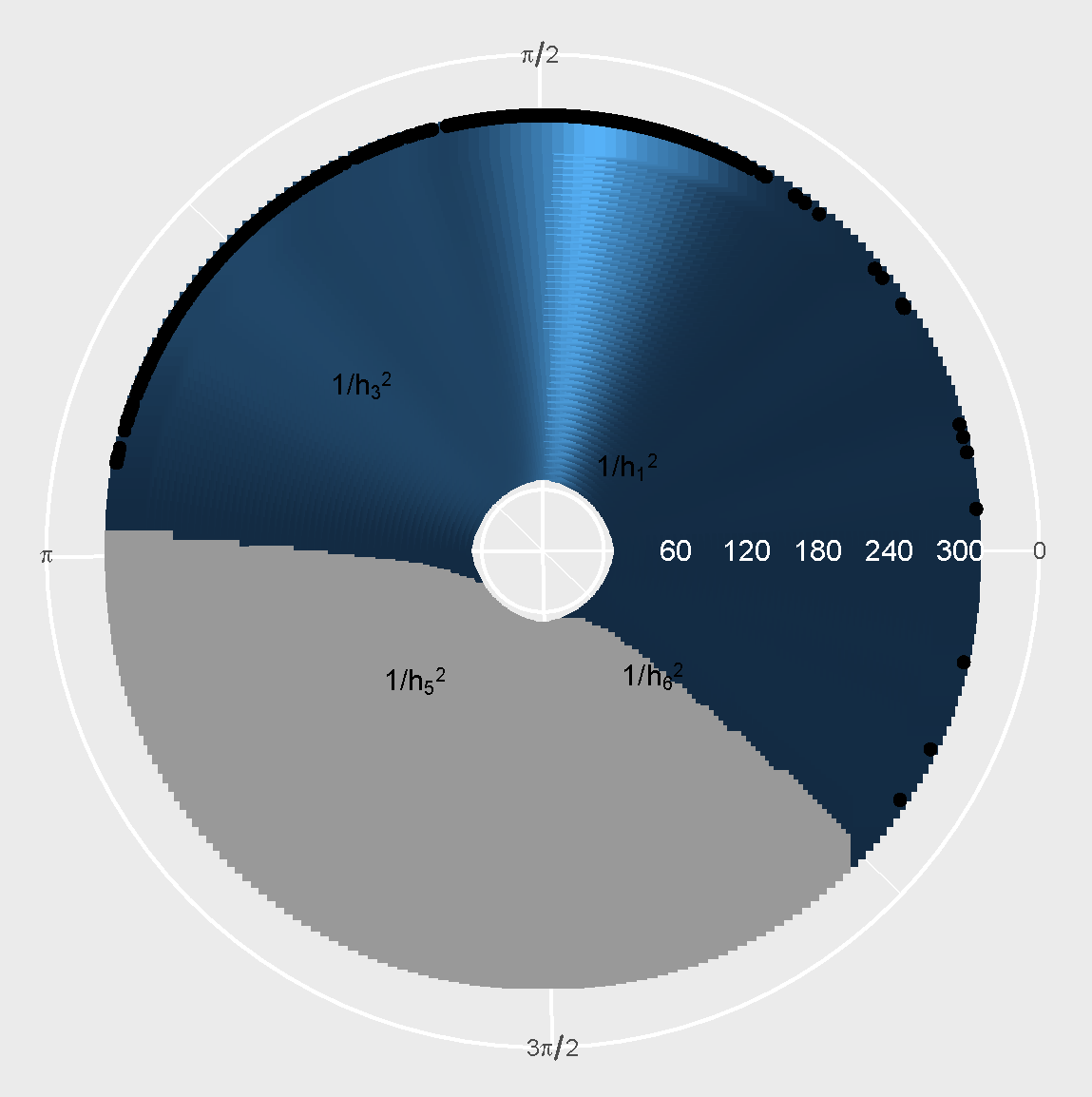}}
        \put(100,175){\includegraphics[scale=.55]{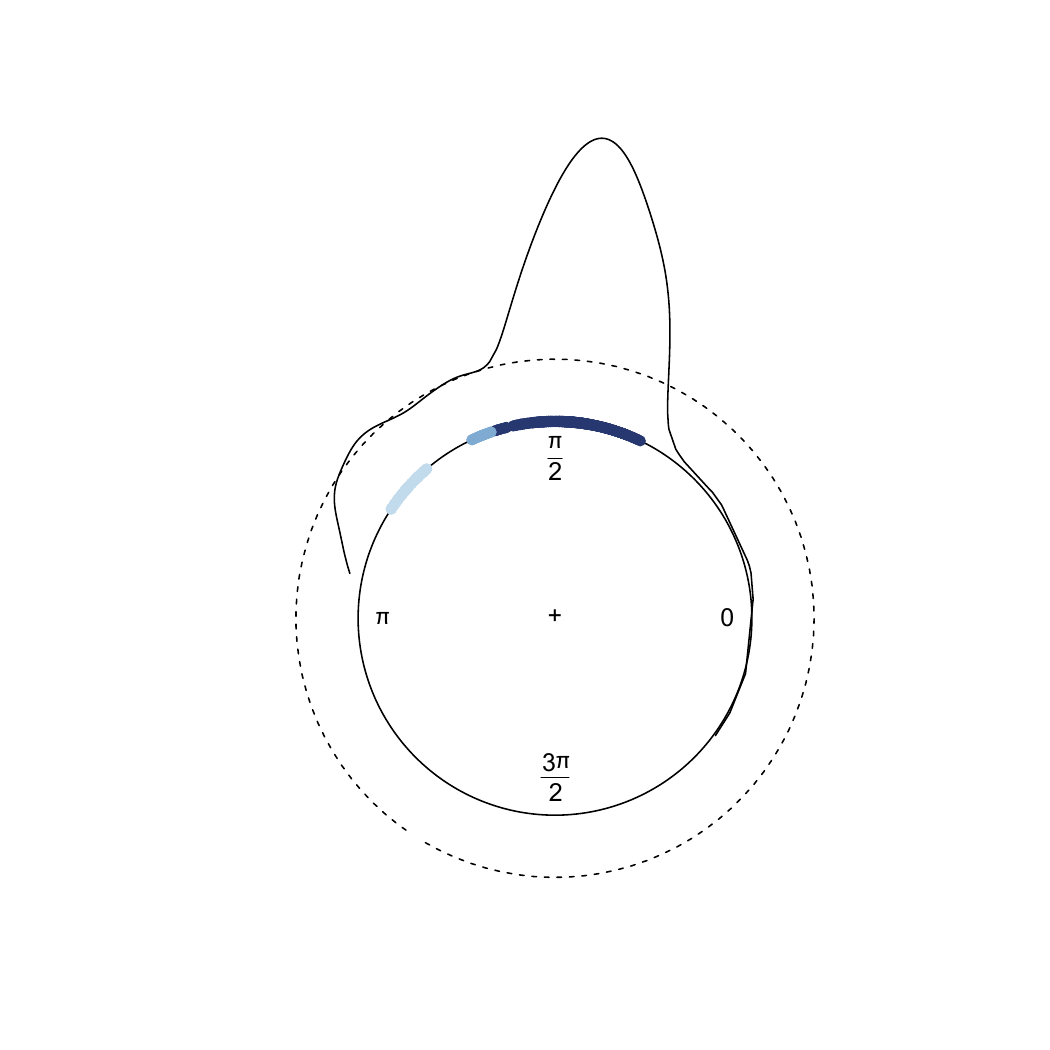}}
        \put(295,188){\includegraphics[scale=.38]{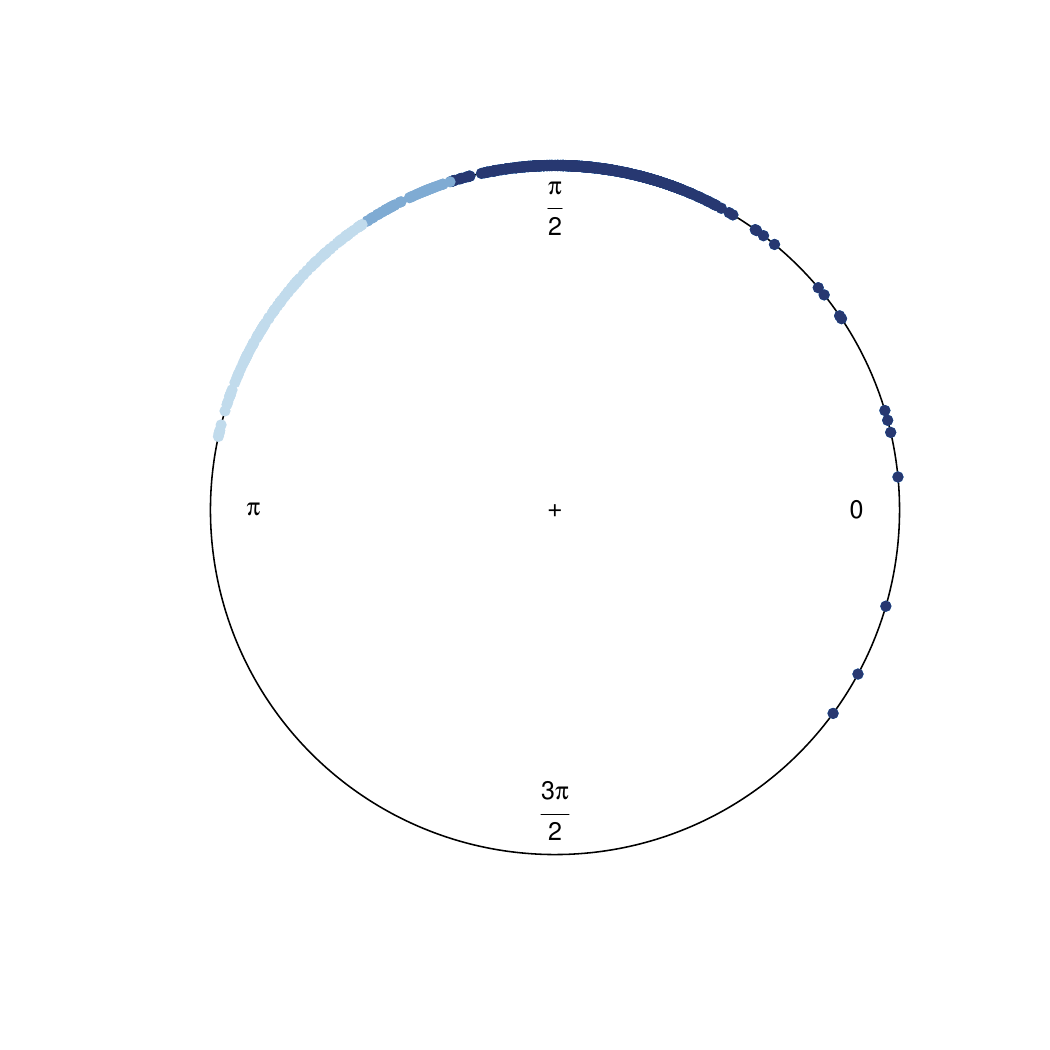}}

	    \put(-10,-10){\includegraphics[scale=.4]{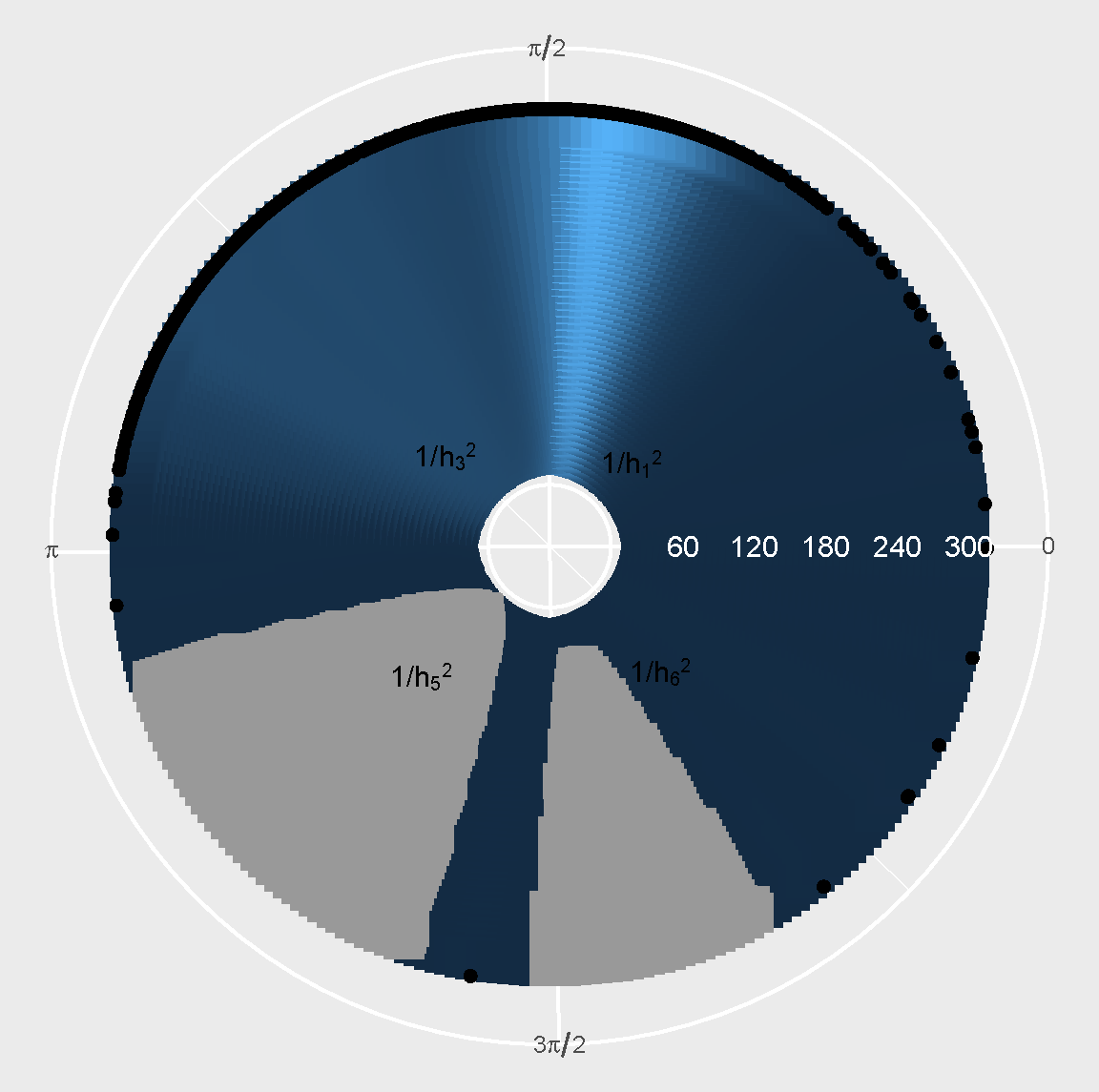}}
	    \put(100,-40){\includegraphics[scale=.55]{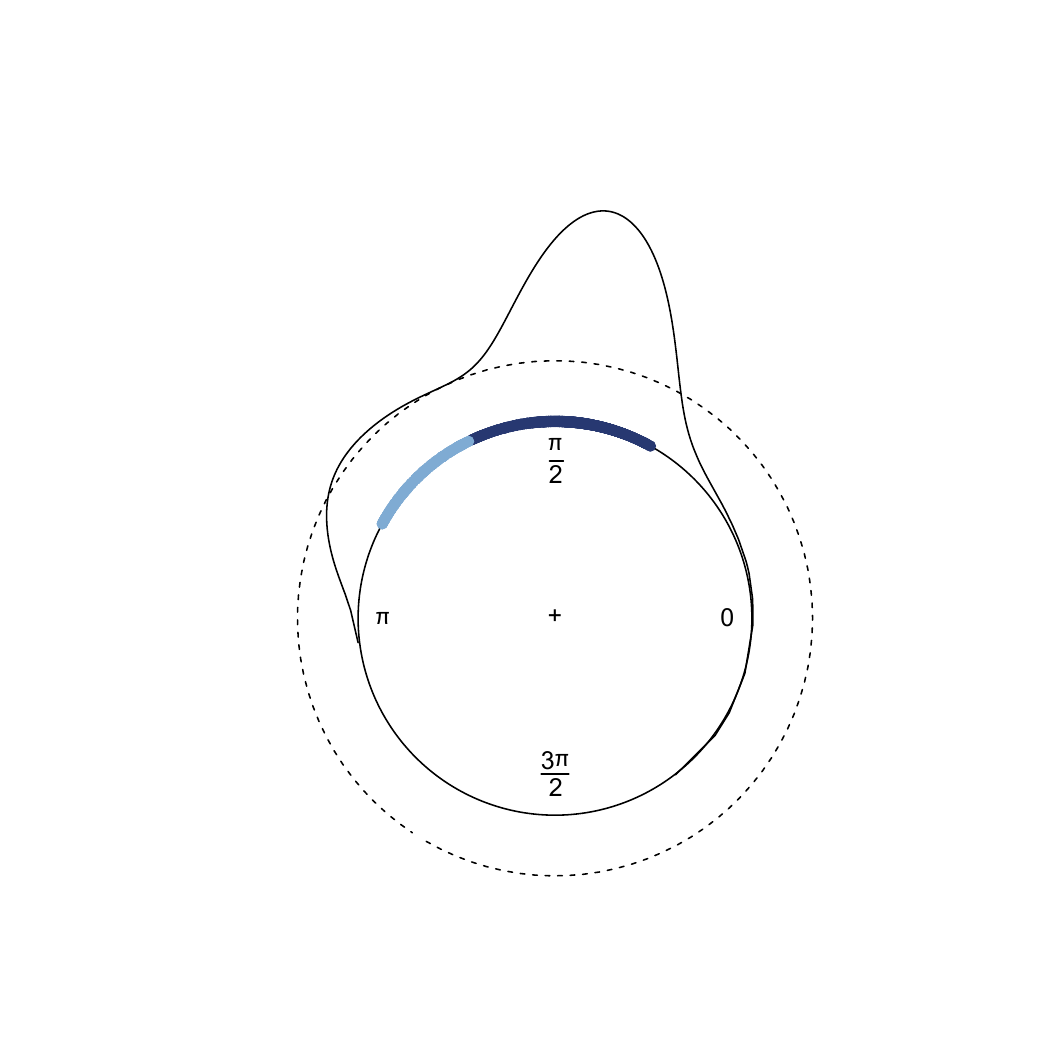}}
	    \put(295,-23){\includegraphics[scale=.38]{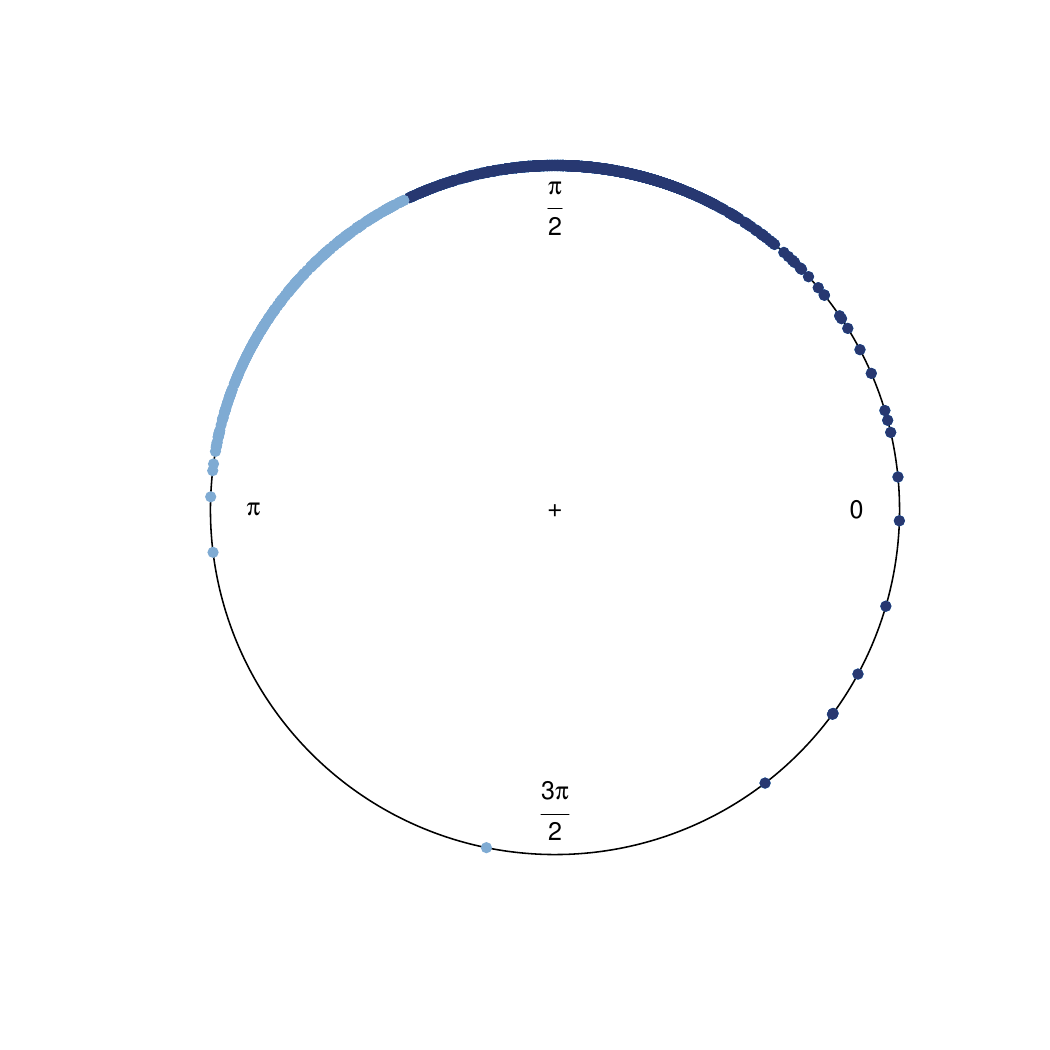}}
    \end{picture} \vspace{0.8cm}
	\caption{In the first row, C\textbf{c}luster for exoplanets discovered in 2014 or before (left), kernel density estimation from cross validation bandwidth ($h_3$) with cluster cores when $1-\tau=0.81$ (center) and final classification (right). In the second row, C\textbf{c}luster for exoplanets discovered in 2021 or before (left), kernel density estimation from cross validation bandwidth ($h_3$) with cluster cores when $1-\tau=0.87$ (center) and final classification (right). }\label{fig:sizerexo}
\end{figure*}

Table \ref{tab:exos1} shows the results of clustering in $S^1$ performed for the values of $1-\tau$ represented in Figure \ref{fig:exo} (left and center). Specifically, it contains the number of clusters, the centers of cluster cores (means vectors), the number of exoplanets in each cluster cores and, finally, intra cores Pearson correlations between variables $M_p$ and $P$ with the associate $p-$value. For exoplanets discovered in 2014 or before, the value of $1-\tau=0.81$ provides three clusters with a balanced number of exoplanets by avoiding groups that are practically empty. In this particular case, it can be checked that means vectors shown in Table \ref{tab:exos1} have a certain degree of similarity with clusters centers in \cite{hung2015intuitive}. Moreover, (two-sided) significant intra core correlations are observed.  For this choice of $1-\tau$, Figure
 \ref{fig:sizerexo} (first row) also contains the C\textbf{c}luster tool for exoplanets (left), the corresponding kernel density estimation from cross validation bandwidth and the associated cluster cores (center) and the final sample classification (right). As for results in Table \ref{tab:exos1} corresponding to exoplanets discovered in 2021 or before, the value of $1-\tau=0.87$ provides the two most balanced groups. Only one cluster core presents (two-sided) significant 
 and positive intra core correlation. Figure
 \ref{fig:sizerexo} (second row) shows the corresponding C\textbf{c}luster tool for exoplanets (left), the kernel density estimation from cross validation bandwidth with cluster cores (center) and the resulting classification (right).

\begin{table*}
 \centering
 \small{
 \begin{tabular}{cccc}
   \hline
  $1-\tau$& Cluster core & Center $(M_p , a, e, [Fe/H], M_s )$& Members \\
\hline
\multirow{3}*{0.36}&C1&$(8.451 , 1.721 , 0.263,  0.031,  1.321 )$&300 \\
    &C2&$(1.609  ,3.889 , 0.258 , 0.087,  1.172   )$&10\\
    &C3&$(0.064 , 0.155, 0.152 , 0.177,  1.089 )$&3\\

\hline
\multirow{3}*{0.48}&C1&$(7.171 , 2.077,  0.256,  0.039 , 1.290)$&376 \\
&C2&$(0.051 , 0.116 , 0.106,  0.126,  0.935)$&37\\
&C3&$( 1.170 , 0.074 , 0.045,  0.100 , 1.317 )$&2\\
\hline

\multirow{3}*{0.53}&C1&$( 6.858,  2.146  ,0.253,  0.046 , 1.276 )$&399 \\
&C2&$(0.049,  0.119 , 0.106, 0.113 , 0.929)$&49\\
&C3&$(0.963 , 0.059,  0.037,  0.119,  1.167)$&14\\
\hline
 \end{tabular}
 }
 \caption{Classification results in $S^4$ for exoplanets discovered up to 2021.}
 \label{tab:clasS4}
\end{table*}

\begin{figure*}[ht!]
\includegraphics[width=0.33\textwidth]{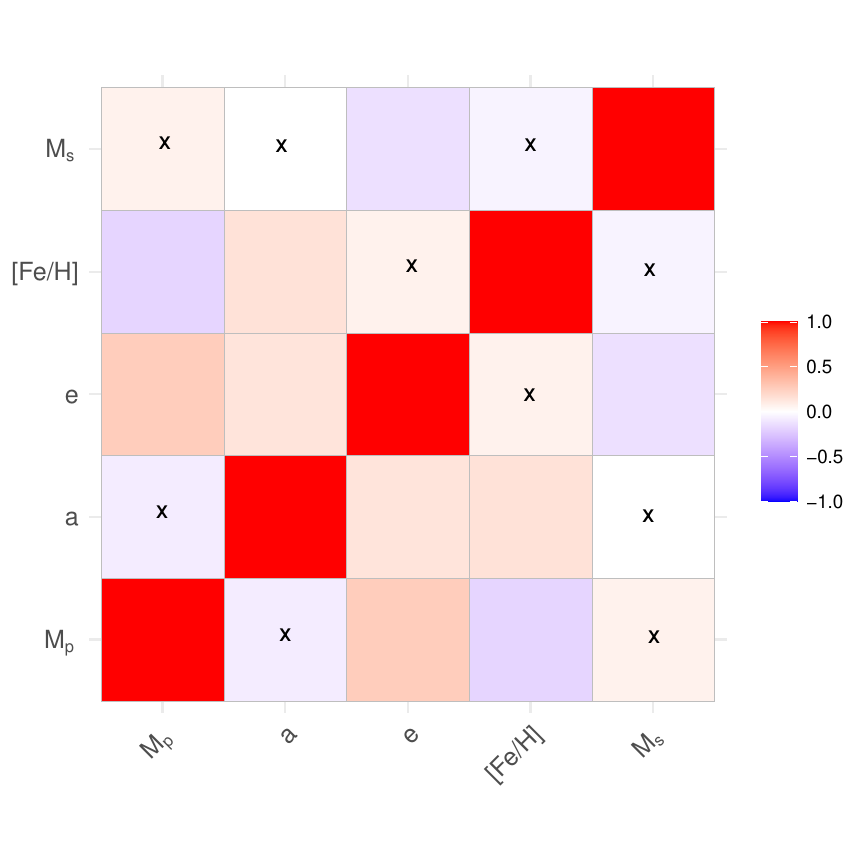}   \includegraphics[width=0.33\textwidth]{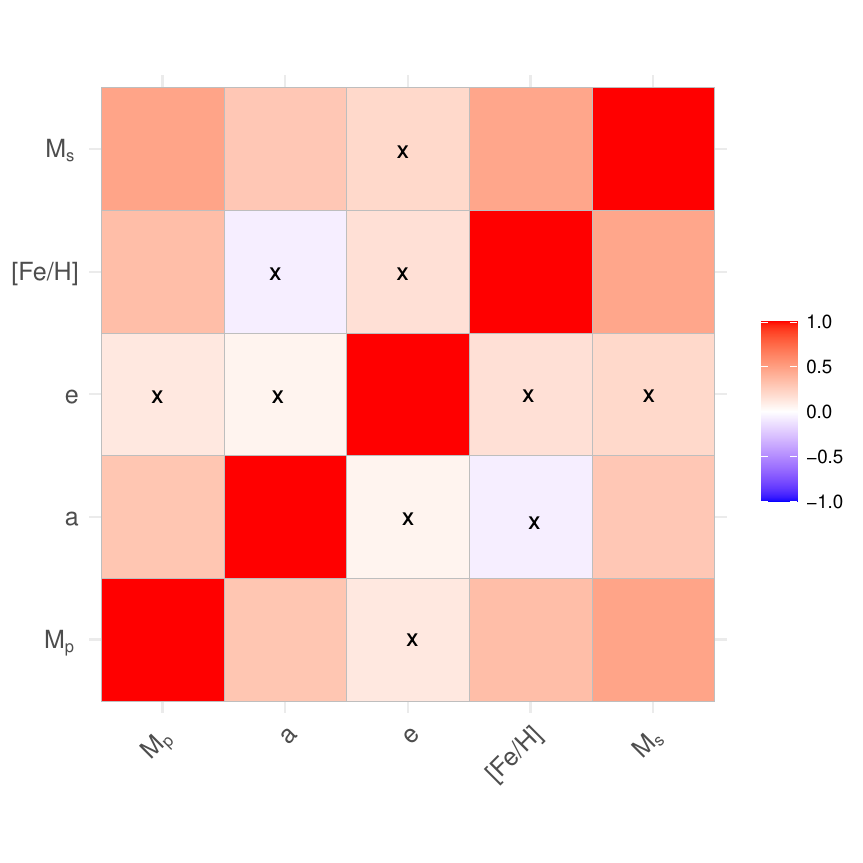}
\includegraphics[width=0.33\textwidth]{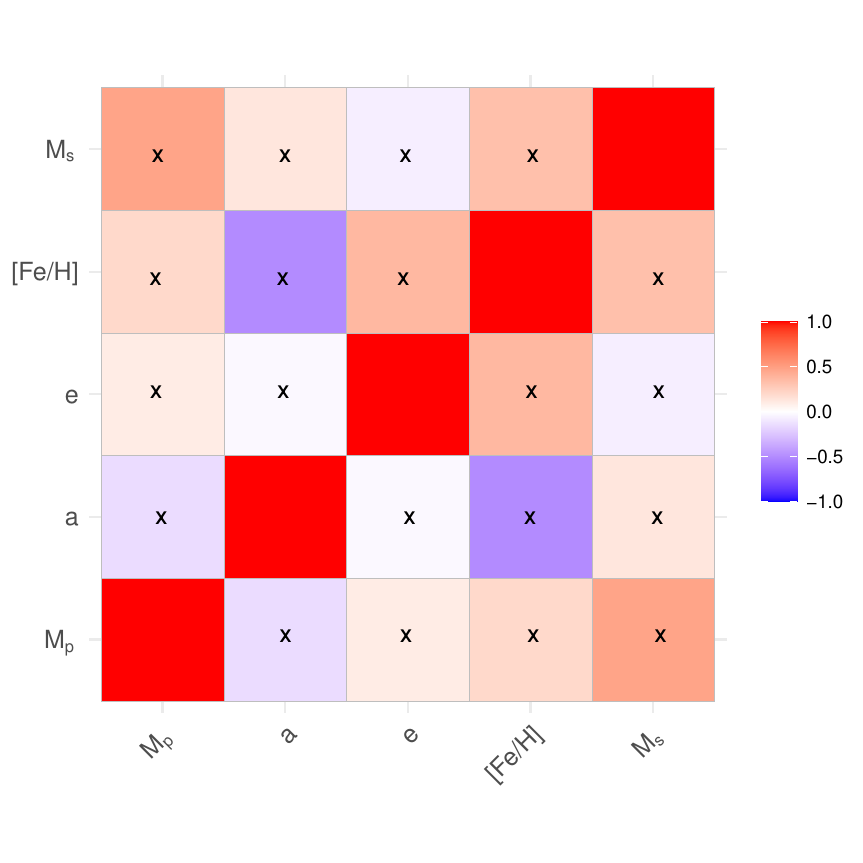}\\
\caption{Intra cluster core Pearson correlation heatmaps of variables $M_p$, $a$, $e$, $[Fe/H]$, $M_s$ involved in clustering analyses performed from density-based method for $S^4$ when $1-\tau= 0.53$ in C1 (left), C2 (center) and C3 (right). Symbol x marks (two-sided) non-significant correlations with significance level $ 0.05$.}\label{Correlations} 
\end{figure*}
 
\begin{figure*}[ht!]
\includegraphics[width=0.33\textwidth]{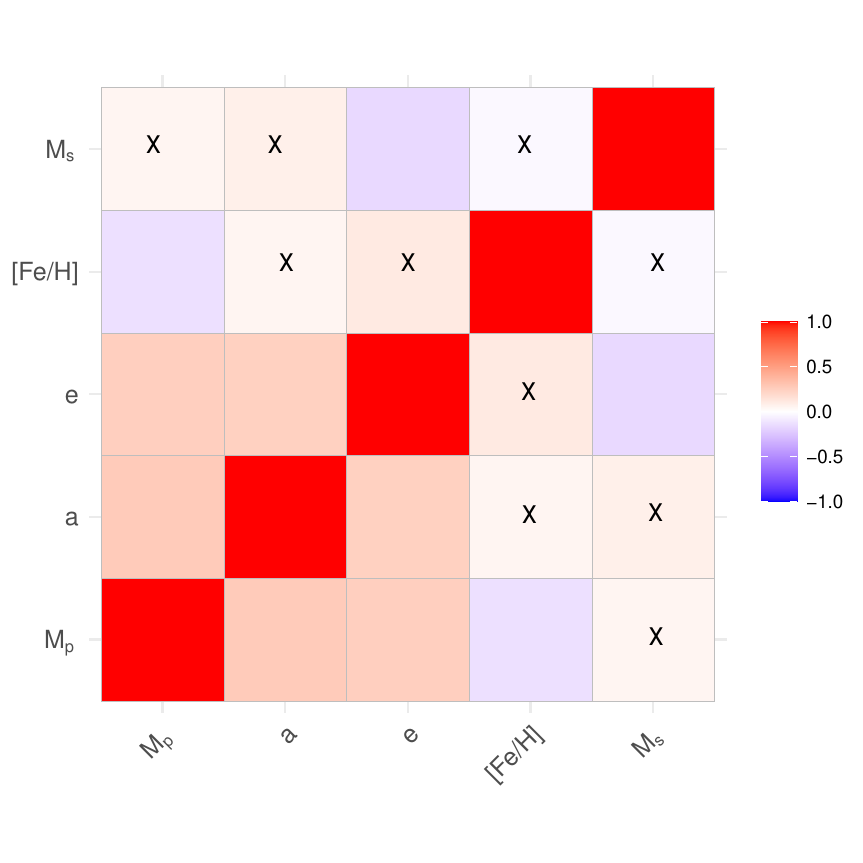}   \includegraphics[width=0.33\textwidth]{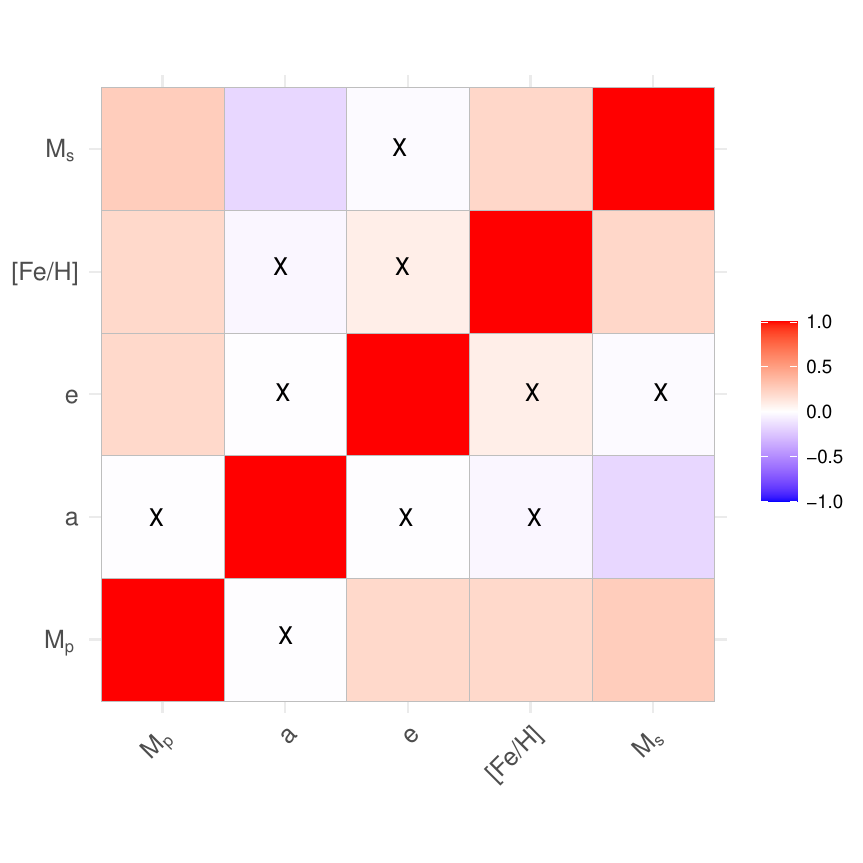}
\includegraphics[width=0.33\textwidth]{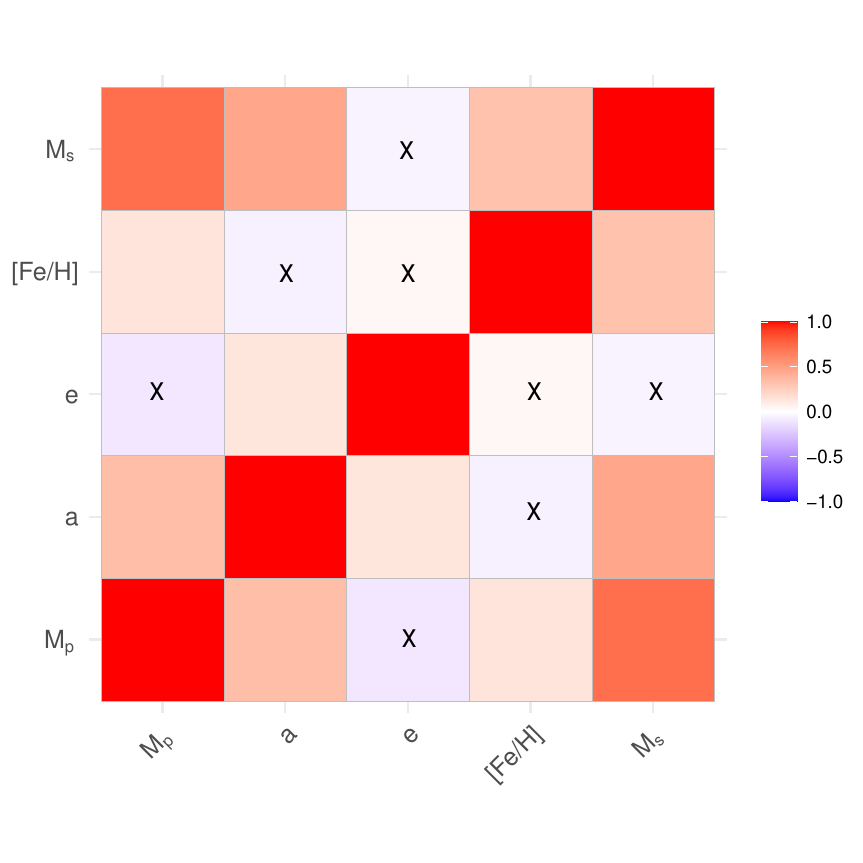}\\
\caption{Intra cluster Pearson correlation heatmaps of variables $M_p$, $a$, $e$, $[Fe/H]$, $M_s$ involved in clustering analyses performed from $3-$means method for $S^4$ in cluster 1 (left), cluster 2 (center) and cluster 3 (right). Symbol x marks (two-sided) non-significant correlations with significance level $ 0.05$.}\label{Correlations3means} 
\end{figure*}

As for clustering analysis in $S^4$, database contains a total of $833$ complete observations on exoplanets discovered in 2021 or before. 
Figure \ref{fig:exo} (right) presents the empirical mode function obtained from kernel density estimation by using a rule-of-thumb for bandwidth selection. Black vertical lines correspond to the different values of $1-\tau$ that will be considered for establishing the cluster cores. Remark that a maximum of three groups of exoplanets are identified in 2021.

Table \ref{tab:clasS4} shows the results of density-based clustering in $S^4$ performed for the values of $1-\tau$ represented in Figure \ref{fig:exo} (right). It contains the cluster core centers (means vectors) and the number of exoplanets in each cluster core. Following \cite{ref1}, it could be analysed which cluster centers are within the regime in which the tidal interaction with the central star. 

Figure \ref{Correlations} contains the Pearson intra core correlation heatmaps of variables $M_p$, $a$, $e$, $[Fe/H]$, $M_s$ when $1-\tau=0.53$. In this case, cluster core C1 contains a total of 399 exoplanets. The significant intra cluster core correlations are: $M_p-e$, $M_p-[Fe/H]$, $a-e$, $a-[Fe/H]$ and $e-M_s$. Note that $M_p$ is positively correlated with $e$. This result implies that higher projected mass exoplanets have higher $e$, thus, the mechanisms for the pumping-up of the eccentricity are more active in high-mass exoplanets for this group. For instance, this is the case of exoplanets HD 122562 b, HD 217850 b, HD 77065 b or MARVELS-16 b. Additionally, $M_p$ is anticorrelated with $[Fe/H]$. This is because of some very massive planets such as  HD 134113 b, HD 283668 b or HD 77065 b present considerable negative values of variable $[Fe/H]$. As for positive correlation between $a$ and $e$, it could indicate that migration levels are smaller as $e$ increases. The highest values of both variables are registered for exoplanets HAT-P-11 c, HD 120084 b, HD 122562 b, HD 211847 b, HD 217850 b, HD 219077 b, HD 219828 c, HD 67087 c or WASP-53 c. Besides, the positive correlation between $a$ and $[Fe/H]$ could imply that the planetary migration is more pronounced for negative values of the stellar metallicity. For example, exoplantes BD+03 2562 b, BD+20 2457 b, HD 11755 b, HD 134113 b, HD 47536 b or  HD 4760 b are under this situation. Finally, $e$ is anticorrelated with $M_s$ in C1. This result implies that there exist exoplanets with big (small) stellar masses and small (big) values of $e$ in exoplanets such as HD 119445 b, nu Oph b or  nu Oph c (HD 108341 A b, HD 22781 b or WASP-53 c).

Cluster cores C2 contains a total of 49 exoplanets. The significant intra cluster core correlations are: $M_p-a$, $M_p-[Fe/H]$, $M_p-M_s$, $a-M_s$ and $[Fe/H]-M_s$.  Since all pairs of variables present positive Pearson correlation coefficients, the existence of direct relationships between them is checked. Unlike C1, $M_p$ is positively correlated with $[Fe/H]$ but also with $a$. As consequence, planets such as HD 102117 b with big values of projected mass also present a high degree of metallicity. Furthermore, exoplanets’ projected masses are positively correlated with stellar masses. This seems natural since higher stellar mass correspond to larger protoplanetary disk surface density, and therefore larger values of $M_p$. This also justifies the existence of correlations between $[Fe/H]-M_s$ and $a-M_s$ in C2.

As regards cluster core C3, it contains a total of 14 exoplanets with non-significant intra cluster core correlations between none of the selected pairs of variables. Specifically, they contain the following planets: CoRoT-29 b, HD 102956 b, HD 13908 b, HD 143105 b, HD 159243 b,
 HD 179949 b, HD 330075 b, K2-29 b,    WASP-108 b,  WASP-119 b, 
WASP-121 b,  WASP-123 b,  WASP-129 b and WASP-84 b.  As in C2, the particular lack of significant correlation between $[Fe/H]$ and $a$ indicates that the stellar metallicity does not play a key role in exoplanet migration for exoplanets in this group.

 For comparative purposes, $3-$means algorithm was also applied on exoplanets dataset in $S^4$. Under this approach, the first cluster identified contains a total of 322 exoplanets and its cluster center is $(8.156,  1.549,  0.263,  0.022,  1.282 )$; a total of 239 observations belong to the second group with center $ (1.595 , 4.629 , 0.250 , 0.065,  1.074) $; and the third cluster is composed by 272 exoplanets and its center is $(0.252,  0.180,  0.133,  0.011 , 0.866 )$. Figure \ref{Correlations3means} contains the Pearson intra cluster correlation heatmaps of variables $M_p$, $a$, $e$, $[Fe/H]$, $M_s$ for the three clusters detected. Comparison of Figures \ref{Correlations} (left) and Figure \ref{Correlations3means} (left) reveals certain similarities between C1 and the first cluster of $3-$means. In particular, common significant correlations are: $M_p-e$, $M_p-[Fe/H]$,  $a-e$ and $e-M_s$. Analysis of Figures \ref{Correlations} (center) and Figure \ref{Correlations3means} (center) shows that $M_p-[Fe/H]$, $M_p-M_s$,  $a-M_s$ and $[Fe/H]-M_s$ correspond to the significant correlations in the second groups for both methods. But, in this case, the sign of Pearson coefficient associate to the pair $a-M_s$ is opposite.

\section{Conclusions and discussion}\label{sec:conclusions}

The main goals of this work are to extend the density-based clustering approach for directional data and to check its classification practical performance. The route designed to reach these objectives can be summarized as follows: (1) Establishing the definition of cluster in \cite{hartigan1975clustering} for data on the unit hypersphere, (2) defining the (population and empirical) mode function and the corresponding cluster tree by solving the associated computational problems, (4) proposing an exploratory tool for analysing the effect of the bandwidth on clustering when kernel density estimation is considered (3) studying the practical behavior of the resulting classification method through simulations and (5) applying directional density-based methods to a real dataset on exoplanets.

Some further research on this topic and some natural extensions are discussed. Firstly, the quality of clusters obtained from density-based methods could be evaluated by adapting the Silhouette information in \cite{menardi2011density} for directional data. Secondly, the consideration of the kernel density estimates proposed in \cite{di2011kernel} (torus) 
and \cite{garcia2013kernel2} (cylinder) enables the adaptation of our proposal to these settings. Furthermore, an R package containing the directional methodology developed in this paper could be implemented as in \cite{azzalini2014pdfcluster} for Euclidean data.

\bmhead{Supplementary material} It contains the s\textbf{C}luster tool.

\bmhead{Acknowledgments} Authors thank Elena V\'azquez Abal and Rosa M. Crujeiras for their help, Giovanni Porzio for providing the exoplanets data and the computational resources of the CESGA Supercomputing Center.

\section*{Declarations}
 P. Saavedra-Nieves acknowledges the financial support of the Xunta de Galicia through the European Regional Development Fund (Grupos de Referencia Competitiva ED431C 2021/24) and of the Spanish Ministry of Science and Innovation through projects PID2020-118101GB-I00 and PID2020-116587GB-I00.


%

\bibliographystyle{bst/sn-basic}
\bibliography{sn-bibliography}


\end{document}


\pagestyle{empty}
\thispagestyle{empty}

\section*{Supplementary material}
\section*{Spherical cluster exploratory tool for bandwidth selection}

 \begin{figure}[h!]
	\begin{picture}(-200,410)
	
	\put(0,157){\animategraphics[width=13.3cm,height=9.2cm]{1}{sizer_}{1}{28}}

		\put(345,220){\mediabutton[
	jsaction={
		if(anim[?taylor?].isPlaying)
	anim[?taylor?].pause();
		else
		anim[?taylor?].playFwd();
		}
		]{\fbox{Play/Pause}}}  
 \end{picture}  \vspace{-4.8cm}
	\caption{s\textbf{C}luster for the spherical sample already considered in Figure 4. In this case, $h_3=0.14$, $h_4=0.13$, $h_7=0.21$, $h_8=0.12$ and $h_9=0.13$ were included in the animation. }\label{fig:cluster}
\end{figure}